\documentclass[10pt,conference]{IEEEtran}
\IEEEoverridecommandlockouts
\usepackage[ruled,vlined,lined,commentsnumbered]{algorithm2e}
\usepackage{booktabs}
\usepackage{balance}
\usepackage{moreverb}
\usepackage{fontenc}
\usepackage{amsmath}

\usepackage{amssymb}
\usepackage{fancybox}
\usepackage{color}
\usepackage{colortbl}
\usepackage{array}
\usepackage{diagbox}
\usepackage{flushend}
\usepackage{multirow}
\usepackage{multicol}
\usepackage{makecell}
\usepackage{graphicx}
\usepackage{setspace}
\usepackage{soul}
\usepackage[utf8]{inputenc}
\usepackage{fontawesome}
\usepackage{dialogue}
\usepackage{listings}
\usepackage{csvsimple}
\usepackage{longtable}
\usepackage[skip=0pt,labelfont=bf]{caption}
\usepackage[breaklinks]{hyperref}
\usepackage{url}
\usepackage{lscape}
\usepackage{rotating}
\usepackage{tikz}
\usepackage{enumitem}
\usepackage{subcaption}
\usepackage{wrapfig}
\usepackage[most]{tcolorbox}
\makeatletter
\@namedef{ver@lineno.sty}{9999/12/31}
\@namedef{opt@lineno.sty}{}
\makeatother
\usepackage{threeparttable}

\usepackage{marvosym}
\usepackage{pifont}
\usepackage{cite}
\usepackage{multicol,multirow}
\usepackage{amsmath,amssymb,amsfonts}
\usepackage{graphicx}
\usepackage{textcomp}
\usepackage{listings}
\usepackage{xcolor,colortbl}
\definecolor{codegreen}{rgb}{0,0.6,0}
\definecolor{codegray}{rgb}{0.5,0.5,0.5}
\definecolor{codepurple}{rgb}{0.58,0,0.82}
\definecolor{backcolour}{rgb}{0.95,0.95,0.92}
\usepackage{marvosym}
\lstdefinestyle{mystyle}{
    backgroundcolor=\color{backcolour},
    commentstyle=\color{codegreen},
    keywordstyle=\color{magenta},
    numberstyle=\tiny\color{codegray},
    stringstyle=\color{codepurple},
    basicstyle=\ttfamily\footnotesize,
    breakatwhitespace=false,
    breaklines=true,
    captionpos=b,
    keepspaces=true,
    numbers=left,
    numbersep=5pt,
    showspaces=false,
    showstringspaces=false,
    showtabs=false,
    tabsize=2
}

\def\BibTeX{{\rm B\kern-.05em{\sc i\kern-.025em b}\kern-.08em
    T\kern-.1667em\lower.7ex\hbox{E}\kern-.125emX}}

\begin{document}

\title{Model Editing for LLMs4Code: How Far are We?
\thanks{$^\dagger$ These authors contributed equally to this work.} \thanks{\textsuperscript{\Letter} Corresponding authors.}
}

\author{\centering
\IEEEauthorblockN{Xiaopeng Li$^\dagger$, Shangwen Wang$^\dagger$, Shasha Li\textsuperscript{\Letter}, Jun Ma\textsuperscript{\Letter}, Jie Yu, Xiaodong Liu, Jing Wang, Bin Ji, Weimin Zhang}
\IEEEauthorblockA{
College of Computer Science,\\
National University of Defense Technology\\
Changsha, China\\
\{xiaopengli, wangshangwen13, shashali, majun, yj, liuxiaodong, wangjing, jibin\}@nudt.edu.cn
\\
wmzhang104@139.com}

}
\maketitle

\begin{abstract}
Large Language Models for Code (LLMs4Code) have been found to exhibit outstanding performance in the software engineering domain, especially the remarkable performance in coding tasks.
However, even the most advanced LLMs4Code can inevitably contain incorrect or outdated code knowledge. Due to the high cost of training LLMs4Code, it is impractical to re-train the models for fixing these problematic code knowledge.
Model editing is a new technical field for effectively and efficiently correcting erroneous knowledge in LLMs, where various model editing techniques and benchmarks have been proposed recently. 
Despite that, a comprehensive study that thoroughly compares and analyzes the performance of the state-of-the-art model editing techniques for adapting the knowledge within LLMs4Code across various code-related tasks is notably absent. 
To bridge this gap, we perform the first systematic study on applying state-of-the-art model editing approaches to repair the inaccuracy of LLMs4Code.
To that end, we introduce a benchmark named CLMEEval, which consists of two datasets, i.e., CoNaLa-Edit (CNLE) with 21K+ code generation samples and CodeSearchNet-Edit (CSNE) with 16K+ code summarization samples. 
With the help of CLMEEval, we evaluate six advanced model editing techniques on three LLMs4Code: CodeLlama (7B), CodeQwen1.5 (7B), and Stable-Code (3B). 
Our findings include that the external memorization-based GRACE approach achieves the best knowledge editing effectiveness and specificity (the editing does not influence untargeted knowledge), while generalization (whether the editing can generalize to other semantically-identical inputs) is a universal challenge for existing techniques.
Furthermore, building on in-depth case analysis, we introduce an enhanced version of GRACE called A-GRACE, which incorporates contrastive learning to better capture the semantics of the inputs. Results demonstrate that A-GRACE notably enhances generalization while maintaining similar levels of effectiveness and specificity compared to the vanilla GRACE. 
\end{abstract}


\section{Introduction}
Large Language Models (LLMs) have demonstrated their powerful understanding and generating capabilities~\cite{zhao2023survey,Chang2024,wei2022emergent,NEURIPS2022_8bb0d291}, and have been applied to areas such as autonomous agents \cite{wang2024survey}, medicine \cite{thirunavukarasu2023large}, and recommendation system \cite{Acharya23}.
LLMs for code (LLMs4Code), trained on massive code-related datasets \cite{Ali2024}, also show remarkable performance in coding tasks within software engineering \cite{xu2022systematic}, including code generation\cite{jiang2024survey,wang2023natural} and code comment generation\cite{chen2021evaluating,geng2024large}.

However, even the most advanced LLMs4Code can contain outdated and incorrect code knowledge due to the following reasons\cite{dou2024whatswrongcodegenerated}. 
On one hand, the training data for LLMs4Code is limited to a certain period, which implies that LLMs4Code cannot learn about the latest software package characteristics from this data as software is continuously changing \cite{lin2022predictive}.
On the other hand, the massive training data inevitably contains some noise\cite{Ji2023}, which ultimately leads LLMs4Code to learn some incorrect code knowledge.
If the aforementioned concerns are not promptly addressed, LLMs4Code will continue to produce bugs or vulnerabilities in production environments\cite{gu2024neuronlevel,tambon2024bugslargelanguagemodels}. A natural way to fix these issues is retraining, but this ``Killing a fly with a cannon'' approach not only consumes a considerable amount of computational resources but also takes a lot of time.
Recently, researchers have explored {\bf Model Editing} to repair LLMs' outdated and incorrect knowledge, whose aim is to efficiently and effectively update targeted knowledge without affecting the other non-targeted knowledge of LLMs\cite{yao-etal-2023-editing}. Existing model editing approaches can generally be categorized into three classes, which are based on External Memorization, Global Optimization, and Local Modification, respectively \cite{wang2023knowledge}.
External Memorization adds an extra module to the original model to store updated knowledge; 
Global Optimization directly or indirectly updates model weights using fine-tuning gradients; 
Local Modification first identifies critical modules in the model where the target knowledge is stored, then uses specific algorithms to update the weights in these key modules. 

State-of-the-art model editing approaches have achieved remarkable success for updating the knowledge of LLMs\cite{zhang2024comprehensive, meng2022massediting, mitchell2022fast, li2023pmet}. Given that, researchers have also explored to utilize model editing to fix errors produced during the code generation process \cite{gu2024neuronlevel}. Despite that, this initial study has certain limitations due to the following reasons. First, tasks in the software engineering domain like code generation and code summarization often involve sequence generations, but the proposed approach by Gu et al.\cite{gu2024neuronlevel} focuses on a single token at each time, which is hard to apply to sequence generation and thus misaligns with the application scenarios. 
Second, this existing study mainly focuses on assessing the ability of the edited model to generate correct contents, while other essential properties of LLMs, such as the fluency, i.e., measuring the extent to which the edited model can generate natural and non-repetitive contents, are ignored.
Third, the existing study focuses on a target model with a maximum of 350M parameters, a scale that is relatively restricted compared to the typical size of general LLMs4Code models, which commonly comprise several billion parameters\cite{zhao2023survey}.
Consequently, the literature has limited understanding regarding the strengths and weaknesses of existing model editing techniques within the context of LLMs4Code, and it remains an open question as to how effectively these techniques can update the code knowledge embedded in LLMs4Code.
There is thus an urge need for a comprehensive empirical study comparing and analyzing the performance of all the state-of-the-art model editing techniques on LLMs4Code. Such a study is necessary and essential, as it can provide answers to fundamental questions such as which types of approaches are the most proficient. This insight can serve as a guideline for researchers in devising more proficient techniques in the future.

To bridge this gap, we perform the first systematic study on applying state-of-the-art model editing approaches to repair the inaccuracy of LLMs4Code. 
To that end, we first build an evaluation benchmark, Code Large Language Models Editing Evaluation (CLMEEval), which consists of two datasets, CoNaLa-Edit (CNLE) and CodeSearchNet-Edit (CSNE), corresponding to the editing of code knowledge in the context of two widely-studied software engineering tasks: a natural language to programming language (NL2PL) code generation task and a programming language to natural language (PL2NL) code summarization task.
Drawing from this benchmark, we employ model editing techniques to rectify the inaccuracies produced by LLMs4Code, mirroring real-world scenarios where updates to code knowledge within a model are necessary. This could involve situations like changes in required APIs for completing specific coding tasks or shifts in the primary functionality of a method due to code changes.
Following the common practice in the model editing domain \cite{Meng2022Locating}, our study evaluates the approaches from four dimensions: {\bf Effectiveness}, the success rate on editing instances; {\bf Generalization}, the success rate on tests that are semantically identical to the editing instances; {\bf Specificity}, the success rate on tests unrelated to the editing instances; and {\bf Fluency}, the fluency of the contents generated by the model \cite{Meng2022Locating}. 
We select six state-of-the-art model editing approaches from the three categories mentioned above and three widely-used LLMs4Code, i.e., CodeLlama (7B)\cite{codellama}, CodeQwen-1.5 (7B) \cite{qwen}, and StableCode (3B)\cite{stable-code-3b} as our study subjects.
Through an extensive evaluation, our study makes the following important findings:

\begin{itemize}
    \item[F1:] The External Memorization-based technique, GRACE, can consistently achieve the optimal effectiveness and specificity across different datasets and LLMs4Code. Nonetheless, all the existing model editing techniques perform poorly in terms of generalization.
    \item[F2:] Most model editing techniques perform comparatively poorly on LLMs4Code, being far less proficient compared with editing general LLMs.
    \item[F3:] Model editing techniques are sensitive to the specific tasks, with all the editing techniques performing worse in NL2PL editing than in PL2NL editing.
\end{itemize}

Moreover, through a case analysis on the best-performing editing approach in terms of the effectiveness and specificity, i.e., GRACE \cite{NEURIPS2023_95b6e2ff}, we identify its weakness in the inability to distinguish between the semantics of different inputs. Based on this observation, we propose an augmented strategy (named as {\bf A-GRACE}) where we introduce an encoder to GRACE that allows it to better capture the input semantics via contrastive learning, and thus improves the generalization of GRACE. 
Results show that A-GRACE significantly improves the generalization while achieves similar effectiveness and specificity compared to GRACE. For instance, on CNLE, A-GRACE improves the generalization of GRACE from almost zero to an average of 76.63\% in terms of the Exact Match metric.

To summarise, our contributions are as follows:
\begin{itemize}
\item {\bf Benchmark.} We construct CLMEEval for LLMs4Code editing, which includes a CNLE dataset with 21K+ NL2PL samples and 16K+ PL2NL samples.
\item {\bf Evaluation.} We evaluate six state-of-the-art model editing approaches on three LLMs4Code, and find that existing model editing approaches can hardly adapt well to LLMs4Code: they usually fail to achieve a good balance among effectiveness, generalization, and specificity.
\item {\bf Strategy.} We propose a refined editing approach A-GRACE, and the experiment results show that A-GRACE achieves promising performances on CLMEEval. Particularly, its generalization is improved by an order of magnitude compared to the vanilla GRACE.
\end{itemize}

\section{Task Definition}
\label{sec:task_def}

The code knowledge embedded in LLMs4Code may be outdated or inaccurate. 
On one hand, software undergoes continuous changes with the addition of new features, bug fixes, and performance enhancements \cite{tao2012software}. As a result, the data used to train LLMs4Code may not accurately capture the most up-to-date code-related knowledge. 
An illustrative instance is the challenge LLMs4Code face in selecting the correct library APIs during code generation tasks \cite{ma2024compositional}.
On the other hand, most LLMs4Code models, including CodeLlama, are trained on data collected from numerous open-source projects. This inevitably introduces a significant amount of noisy information into the training data, which can potentially result in incorrect outputs from LLMs4Code models \cite{Ji2023}.
Therefore, it is necessary to regularly update the code knowledge of the LLMs4Code.
Our study investigates the proficiency of model editing approaches on achieving such a target. In the following, we provide the formal definition for this task.

Denote an LLMs4code as $f$ and an editing instance as a tuple $M=(X,Y,\mathbb{C},\mathbb{U})$. Here, $X$ and $Y$ are the input and label of the editing instance; $\mathbb{C} = \{X^{\prime}_1,...\}$ is the set of inputs where each $X^{\prime} \in \mathbb{C}$ is semantically identical with $X$; 
$\mathbb{U}=\{U_1,...\}$ is the set of instances unrelated to the editing instances where for each $U=(X_u, f(X_u)) \in \mathbb{U}$, $X_u$ is semantically inconsistent with $X$ and $f(X_u)$ is the corresponding output of the LLMs4Code. 
The objective of model editing in LLMs4code is to enable LLMs4code to accurately identify editing instances and produce corresponding outputs \cite{yao-etal-2023-editing}. 
In other words, the edited model is capable of generating desired outputs for the targeted inputs, while not affecting other non-targeted inputs or the original capabilities of the model.
Formally, denote the edited LLMs4code as $f_e$,   
for each editing instance $M$, $f_e$ should satisfy the following conditions:
\begin{align}
    &f_e(X) = Y \label{equ:td1}\\
    &f_e(X^{\prime}) = Y, \forall X^{\prime}\in \mathbb{C}  \label{equ:td2}\\
    &f_e(X_u) = f(X_u), \forall U\in \mathbb{U}  \label{equ:td3}
\end{align}
\eqref{equ:td1} means model editing is capable to generate desired outputs;
\eqref{equ:td2} means that the edited model $f_e$ can generalize well to semantically identical inputs;
and \eqref{equ:td3} means that model editing should not misidentify other non-targeted instances, thereby avoiding affecting the original knowledge of the model.

\section{Study Design}

\subsection{Model Editing Techniques Selection}
\label{subsec:me_selection}

Model editing, a rapidly developing field with various techniques having been proposed successively \cite{zhang2024comprehensive}, can be divided into three categories based on the source of new information and the method of updating the information during the editing process: External Memorization, Global Optimization, and Local Modification \cite{wang2023knowledge}. For each category, we select the most representative and state-of-the-art techniques as our study subjects, based on the introduction from a recent survey \cite{wang2023knowledge}.


Overall, our study takes totally six model editing techniques into consideration, including one External-Memorization-based approach, two Global-Optimization-based approaches, and three Local-Modification-based approaches.
Table \ref{tab:Selected_Techniques} lists the categorized selected techniques and the following briefly introduce the working mechanism for each of them. Due to the space constraint, readers can refer to the original papers for more detailed information.

\begin{table}[t]
\caption{Selected techniques in this study.}
\scalebox{0.75}{
\begin{tabular}{c|c|c}
\hline
\diagbox{Categorization}{\# Edited layers} & Single layer & Multiple layers \\ \hline
Global-Optimization & FT-L\cite{zhu2020modifyingmemoriestransformermodels} & MALMEN\cite{tan2024massiveeditinglargelanguage} \\ \hline
Local-Modification & ROME\cite{Meng2022Locating} & MEMIT\cite{meng2022massediting}, PMET\cite{li2023pmet} \\ \hline
External-Memorization & GRACE\cite{NEURIPS2023_95b6e2ff} & - \\ \hline
\end{tabular}}
\label{tab:Selected_Techniques}
\end{table}

\textbf{GRACE} \cite{NEURIPS2023_95b6e2ff}: This method is an External Memorization approach. The primary difference between this type of approaches and the other two categories is that these techniques do not directly update model weights but rather add modules to memorize new knowledge. 
That is to say, this type of method maintains the original weights of the model at the cost of longer inference time.
GRACE is an advanced approach in this category, creating a discrete local codebook of mapping key-values in the hidden space of the model and designing a deferral mechanism to enable the codebook to recognize the key-value pairs of edited knowledge.

\textbf{FT-L} \cite{zhu2020modifyingmemoriestransformermodels}: This method is a Global Optimization approach, which fine-tunes the model directly on the editing dataset while employing an explicit parameter space $L_{\infty}$ norm constraint to minimize the impact on non-targeted data as much as possible.
Given that the volume of data for editing is usually small, this method fine-tunes only a small part of the parameters of the model (e.g., parameters from one layer).

\textbf{MALMEN} \cite{tan2024massiveeditinglargelanguage}: This method employs Global Optimization. Unlike FT-L, which directly updates weights based on backward-propagation gradients, MALMEN post-processes such gradient information through a hyper-network to calculate a theoretical offset (named as the {\em parameter shift}) for each parameter, and then performs parameter updates.

\textbf{ROME} \cite{Meng2022Locating}: This method falls under the Local Modification category. It first uses causal intervention \cite{NEURIPS2020_92650b2e} to identify a single key layer where factual knowledge is stored in LLMs. ROME assumes that the Feed Forward Network (FFN) of the key layer has a key-value mapping capable of recalling factual knowledge \cite{geva-etal-2021-transformer}, models the editing task as a constrained least squares problem \cite{Meng2022Locating}, and then utilizes the closed-form solution of the constrained least squares for Rank-One model editing (i.e., it only supports editing a single piece of knowledge at a time).

\textbf{MEMIT} \cite{meng2022massediting}: This method is also a Local Modification approach. Similar to ROME, it views the FFN as a key-value mapping and updates the key layers identified by causal intervention. However, unlike ROME, MEMIT updates the weights of multiple key layers. 

\textbf{PMET} \cite{li2023pmet}: The motivation of PMET is that the hidden states of the whole Transformer layer used by ROME and MEMIT may contain unnecessary or even noisy information unnecessary for the weight update, which can lead to imprecise updates and ultimately affect the editing performance.
Therefore, PMET opts to only focus on the hidden states of the FFN module and exclusively update the weights of FFN for more precise updates.


\subsection{Editing Subjects}
We select three popular open-source LLMs4Code: 
\begin{itemize}
    \item \textbf{CodeLlama (7B)}\cite{codellama}: It is developed by Meta, which is originated from Llama-2 \cite{touvron2023llama2openfoundation} and trained on 0.5 trillion tokens of code data.
    \item \textbf{CodeQwen-1.5 (7B)}\cite{qwen}: It is developed by Alibaba, which is originated from Qwen1.5 \cite{qwen} and trained on 3 trillion tokens of code data.
    \item  \textbf{StableCode (3B)}\cite{stable-code-3b}: It is developed by stability.ai, which is pre-trained on 1.3 trillion tokens of code data.
\end{itemize}
These three LLMs4Code are all composed of 32-layer Transformer decoders.


\subsection{Dataset Construction}
\begin{figure}[t]
    \centering
    \includegraphics[height=3.8cm]{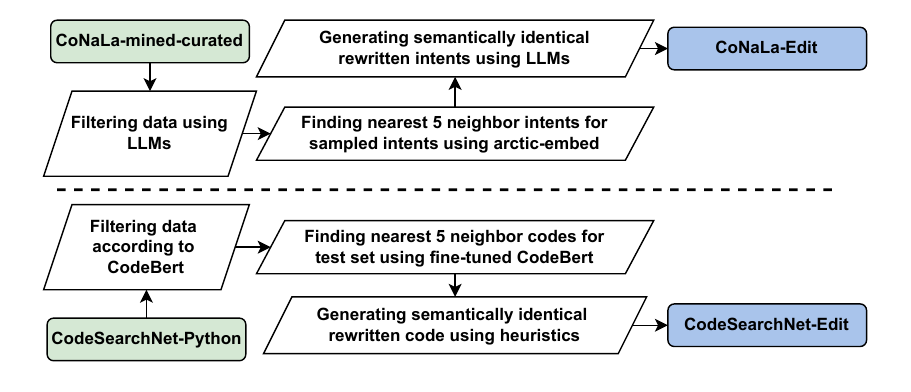}
    \caption{Data construction process of the CNLE and CSNE datasets}
    \label{fig:data_construct_flow}
\end{figure}

To ease the evaluation of our study, we build a benchmark for LLMs4Code editing, named CLMEEval, which considers two types of code-related tasks: NL2PL (the CNLE dataset) and PL2NL (the CSNE dataset). 
We build the CNLE and CSNE datasets based on the existing CoNaLa \cite{conala} and CodeSearchNet \cite{husain2019codesearchnet} datasets, respectively. The CoNaLa dataset was crawled from Stack Overflow, with two different versions (i.e., the {\em curated} and {\em mined} versions) containing 2.88K and 594K pairs of Python code snippets and natural language descriptions (a.k.a, the intents or requirements), respectively. 
To construct a more comprehensive dataset, we rely on the {\em mined} version for data construction. Specifically, we focus on an officially-released version {\em conala-mined-curated}\footnote{\url{https://huggingface.co/datasets/codeparrot/conala-mined-curated}} that further processes the {\em mined} version of CoNaLa, rewriting the intents for the code snippets to improve the clarity.
The CodeSearchNet dataset was collected from GitHub and includes 2 million code-comment pairs across six programming languages.
The overall construction process of the CNLE and CSNE datasets is shown in Figure \ref{fig:data_construct_flow}.

\subsubsection{CoNaLa-Edit (CNLE)}

{\bf Step 1: Filtering.}
It has been observed that natural language descriptions and code snippets may have semantic discrepancies \cite{sun2022importance}, which may bring noise to our evaluation. 
To ensure the reliability of our dataset, we decide to further select high-quality code and natural language pairs from the dataset.
To that end, we use DeepSeek-V2 \cite{deepseekv2} to evaluate the semantic matches between the code and the description.
Specifically, we ask DeepSeek-V2 to evaluate the degrees into five levels: Very Irrelevant, Irrelevant, Neutral, Relevant, and Very Relevant.
To ensure the output of DeepSeek-V2 is steady, we set the temperature to zero.

To verify the reliability of the scoring of DeepSeek-V2, we manually sample 384 pairs out of the 594k pairs for manual validation (confidence level: 95\%, margin of error: 5\%). 
The first two authors label each pair independently, categorizing the matching degree into five levels as mentioned above.
After labeling, the Cohen’s kappa coefficient of agreement \cite{cohen1960coefficient} between the two authors is 0.91, indicating a high degree of consistency.
As for the pairs labeled differently, we hold several meetings to resolve the disagreements until an agreement is reached.
After that, we calculate the Cohen’s kappa coefficient of agreement between the manual labelling and the scoring of DeepSeek-V2.
The result is 0.70, also indicating a high degree of consistency and thus the scoring from DeepSeek-V2 is reliable.
The scoring statistics of DeepSeek-V2 are shown in Table \ref{tab:deepseek_score}. In our study, we select the 170,487 high-quality samples scored as {\em Very Relevant} for further processing. We deduplicate the cases whose intents are identical and finally use the left 126,330 samples for further construction.

\begin{table}[t]
\caption{Statistics of the labeling score of DeepSeek-V2.}
\label{tab:deepseek_score}
\scalebox{0.74}{
\begin{tabular}{cccccc}
\toprule
Level   & Very Irrelevant      & Irrelevant       & Neutral      & Relevant       & Very Relevant       \\ \hline
Numbers & 66,205 & 111,063 & 97,445 & 148,691 & 170,487 \\ \bottomrule
\end{tabular}}
\end{table}

{\bf Step 2: Specificity Evaluation Preparation.}
The specificity evaluates to what extent the editing method affects non-targeted knowledge. That means we need to check if the modification to the model leads to result shifts on a non-targeted input.
In our benchmark, we decide to split the dataset into different non-overlapped parts (i.e., the {\em targeted data} and {\em non-targeted data}).
The targeted data are used to evaluate the effectiveness of model editing, while the non-targeted data are used for the specificity assessment.
The rationale of this decision is that with the new advanced lifelong learning or batch editing techniques \cite{zhang2024comprehensive}, one might be able to update the model knowledge for a number of inputs simultaneously in the future, and our design can safely support the evaluation under such a situation with no overlap data between the targeted and non-targeted data.
Specifically, we randomly sample $21,055$ ($= 126,330/6$) instances as the targeted data, and the remaining ones are treated as non-targeted data. 

In our evaluation towards specificity, we target the most challenging scenario, that is, if the edited model performs differently on the non-targted inputs that are semantically similar to the targeted inputs. 
To identify semantically similar intents, we employ the advanced embedding model arctic-embed \cite{merrick2024arcticembedscalableefficientaccurate} to embed the intents of the descriptions.
Then, for each input description from the targeted data as an editing instance, we match it with the five nearest neighbor samples from the non-targeted data in terms of their embeddings. These selected five samples are used to assess the specificity on this editing instance.
Specifically, we compare if the edited model generates different outputs on such neighbor instances compared with its original output.

{\bf Step 3: Generalization Evaluation Preparation.}
The generalization evaluates to what extent the edited model can success on descriptions that are semantically identical but syntactically different to the editing target. 
Inspired by the powerful generation ability on natural languages of LLMs \cite{zhao2023survey}, we decide to utilize DeepSeek-V2 to generate such rewritten intents for the $21,055$ editing instances.
We enable LLMs to freely rewrite such intents in our aim to create a diversified dataset, rather than relying on pre-defined heuristics.
The prompt we used is demonstrated in our online repository.

\subsubsection{CodeSearchNet-Edit (CSNE)}
{\bf Step 1: Filtering.}
We recall that our CNLE dataset exclusively contains Python code snippets.
To maintain consistency with the $<$Code snippet, Comment$>$ pairs in CNLE, we choose to use the Python subset from the CodeSearchNet. 
The code length in the CodeSearchNet-Python dataset varies significantly, with a number of code snippets exceeding the maximum limit of the state-of-the-art pre-trained code models (e.g., 512 tokens for CodeBERT \cite{feng-etal-2020-codebert}).
Considering that we will rely on pre-trained code models to capture the semantics of various code snippets in the subsequent steps, we have made the decision to filter out code snippets with a token count exceeding 512. This filtering step aims to ensure that we obtain precise semantic representations from CodeBERT.
As a result, we obtain 318,336 samples from the training-validation set of CodeSearchNet-Python and 16,020 samples from the test set of CodeSearchNet-Python, and all samples from the test set are used as editing instances. In other words, cases from the test set are used as the targeted data, while cases from the training-validation set are used as the non-targeted data.

{\bf Step 2: Specificity Evaluation Preparation.}
For each sample in the test set, we match it with five nearest samples from the training-validation set to evaluate specificity.
Specifically, we use CodeBERT, which is fine-tuned on the code search task with the CodeSearchNet-Python dataset, to embed all code snippets in the training-validation and test sets, and then select five samples with the closest cosine similarity from the training-validation set as neighbor samples. 

{\bf Step 3: Generalization Evaluation Preparation.}
The input of code summarization task is a code snippet. Therefore, to evaluate the generalization, we need to create semantic identical but syntactic different code snippets. A natural way is to utilize LLMs, but the existing study has shown that LLMs cannot achieve a perfect accuracy when revising the code \cite{guo2024exploring}. 
On the contrary, we choose to use existing program transformation techniques to ensure the correctness of the generated code.
Specifically, we reproduce the ALERT approach \cite{yang2022natural} where we first identify the variables in the code and replace the variables by the masked prediction from CodeBERT. We generate a variant for each case from the test set, and use these code to evaluate the generalization.

\subsubsection{Dataset statistics}
\label{sec:Dataset_statistics}

\begin{table}[t]
\caption{Dataset statistics of CNLE and CSNE datasets.}
\scalebox{0.72}{
\begin{tabular}{c|ccc|ccc}
\toprule
Dataset           & \multicolumn{3}{c|}{CNLE} & \multicolumn{3}{c}{CSNE} \\ \hline
Statistical value & Mean          & Median  & \# Num  & Mean         & Median  & \# Num \\ \hline
Effectiveness     & 15.36   & 12 &21,055 & 239.17   & 218 & 16,020\\
Generalization    & 19.21   & 17& 21,055 & 200.73 & 183&16,020  \\
Specificity     & 14.14& 12 &105,275 &261.32& 242&80,100 \\ \hline
\makecell{Effectiveness/\\Generalization-target}  & 30.42    & 18 & 21,055& 57.35   & 31 &16,020\\ \hline
Specificity-target     &19.29 & 19 &315,825&20.68&20 &240,300 \\
\bottomrule
\end{tabular}}
\label{tab:data-statistics}
\end{table}

The statistics of the two datasets are shown in Table \ref{tab:data-statistics}.
Lines beginning with Effectiveness, Generalization, and Specificity demonstrate the statistical information of the inputs (requirements for the CNLE dataset and code snippets for the CSNE dataset) to evaluate the respective capacity of model editing techniques. 
While the rest two lines demonstrate the statistical information of the oracle outputs that correspond to different inputs. Note that the oracle outputs for Effectiveness and Generalization are identical, and they are thus merged in the table. 
Furthermore, the oracle output in terms of Specificity is the original output of the specific LLMs4Code being considered, which can vary when different LLMs4Code models are used. Therefore, the total number in the last line is calculated as the number of considered LLMs4Code (3) $\times$ the number of instances to evaluate Specificity (e.g., 315,825 and 240,300). 

\subsection{Evaluation Metrics}
To comprehensively assess the effects of model editing techniques on targeted and non-targeted instances, we evaluate the performance of editing techniques from four aspects, namely,  Effectiveness, Specificity, Generalization, and Fluency.

All of the first three aspects require to compare the output from the model with the oracle (the human-written content): for Effectiveness, the comparison result represents how well the model works on the targeted inputs; for Specificity, the result represents how well the model can distinguish between targeted and non-targeted inputs; for Generalization, the result represents how well the model works on inputs that are semantically identical to the targeted inputs.
Given that, we utilize the same metrics to evaluate the first three aspects, which are {\bf Exact Match} (EM), {\bf BLEU} \cite{BLEU} and {\bf ROUGE-L} \cite{ROUGE}.
EM evaluates whether the output completely matches the reference.
BLEU and ROUGE-L are widely used evaluation metrics in code-related tasks \cite{wan2018improving, bansal2021project}. BLEU measures the degree of n-gram overlap between a sequence and a set of reference sequences; the higher the overlap, the higher the BLEU score. ROUGE-L is the most popular variant of ROUGE, which specifically calculates the longest common subsequence between the output and the reference.

Fluency evaluates the impact of model editing techniques on the generation ability of the model. We follow Meng et al. \cite{Meng2022Locating} in using $-\sum_{k}{f(k)\text{log}_2f(k)}$ to assess fluency, where $f(k)$ is the n-gram frequency distribution of the n-gram $k$ from the generated content.

For all the aforementioned metrics, a higher value indicates better performance.

\subsection{Research Questions}

Our study aims to answer the following questions.

\begin{enumerate}
  \item[(1)] \textbf{RQ1. How proficient are existing model editing approaches at updating LLMs4Code?}
  We first systematically investigate the performance of model editing techniques selected in Section \ref{subsec:me_selection} for editing LLMs4Code. To answer this question, we use the CLMEEval benchmark to evaluate these model editing techniques across 3 LLMs4Code. 
  Specifically, we first use the model editing techniques to adjust the knowledge of LLMs4Code. After that, we test the effectiveness, specificity, and generalizability of the edited model through our prepared dataset. 
  Finally, we use the targeted input and its semantically-identical variant (the data used to evaluate generalization) as model inputs, and allow the  edited LLMs4Code to freely generate 64 tokens to test fluency. The final result is the average fluency of these two inputs.
  When testing effectiveness, specificity, and generalizability, it is crucial to ensure the consistency of generation of LLMs4Code, for which we use a greedy search strategy. 
  As for testing fluency, which assesses the generative capability of the edited model, we set Top-K=5 and use beam search to allow the edited LLMs4Code to generate freely.
  
  \item[(2)] \textbf{RQ2. How efficient are the existing editing techniques for LLMs on LLMs4Code?}
  In addition to the proficiency, editing efficiency is also a focal point of model editing which may influence the developers' adoption in practice. We address this question by evaluating the editing efficiency of model editing techniques on LLMs4Code. 
  For each instance edited by each editing techniques, we record the time and peak memory costs.
  
  \item[(3)] \textbf{RQ3. How can we improve the model editing techniques for better performances on LLMs4Code?}
  Based on the results of RQ1 and RQ2, we further analyze the in-depth reasons behind the performance of existing editing techniques, and attempt to utilize the observations to improve current techniques, aiming to enhance the performance of editing techniques on LLMs4Code.
\end{enumerate}

\subsection{Experiment Settings}
All experiments are conducted on A800 GPUs. We use the basic prompt when asking the LLMs to finish a task. Specifically, on the CNLE dataset, we provide the intent description and ask the LLMs to generate the code (and vice versa on the CSNE dataset).
For the methods that require training, we use leave-one-out cross-validation and divide the targeted data of CNLE and CSNE into a test set
and a training set with a 1:1 split. The scale of test set in this division is consistent with existing model editing benchmarks \cite{cohen-etal-2024-evaluating,zhong-etal-2023-mquake,akyurek-etal-2023-dune}.
We then consolidate the outcomes from the two rounds of training and testing as the final results for the complete dataset.
Previous works have explored the optimal hyperparameters for the selected model editing techniques on the Llama-2 (7B) model \cite{touvron2023llama2openfoundation,wang2024easyediteasytouseknowledgeediting}.
Considering that CodeLlama (7B) shares the same architecture as Llama-2 (7B), we follow the previous works when setting hyperparameters. 
Unless otherwise specified, we set the same hyperparameters for the three LLMs4Code models when evaluating a particular model editing method. 
Additionally, unless otherwise noted, the inputs and labels for these techniques are the $X$ and $Y$ of the edit instance tuples (cf. Section~\ref{sec:task_def}). 
The following details the experiments for each selected model editing technique.

\textbf{FT-L:} This approach was designed to only target single-token editing cases, whereas both our datasets are multi-token editing cases (i.e., the sequence generation). Therefore, we implement FT-L with the standard fine-tuning schema to adapt it to our datasets (i.e, FT-M in \cite{zhang2023large}), setting the down projection matrix of the 21st layer of LLMs4Code as learnable parameters.
It iterates up to 40 times with a learning rate of $5\times 10^{-4}$. We set the norm constraint $\delta=1\times 10^{-4}$ to prevent significant changes in model parameters and to protect the original knowledge of the model.

\textbf{ROME, MEMIT, PMET:} These techniques first use the causal intervention \cite{NEURIPS2020_92650b2e} to locate critical layers storing factual knowledge and then edit those critical layers. 
An existing work shows that there is negligible correlation between the location results from causal intervention and model editing performance.
Therefore, we do not use causal intervention to select critical layers but directly refer to the existing studies \cite{li2024sweaupdatingfactualknowledge,wang2024easyediteasytouseknowledgeediting} which have already identified the critical layers of the Llama-2 (7B) model. 
Specifically, ROME, designed for a single layer editing, edits the 5-th layer of the three LLMs4Code, while MEMIT and PMET edit the \{4,5,6,7,8\}-th layers of the three LLMs4Code. 

\textbf{MALMEN:} MALMEN updates the down projection matrices of the last 5 layers of LLMs4Code. It first trains a hypernetwork on the training set to predict parameter shifts. Specifically, the hypernetwork of MALMEN consists of two consecutive MLP layers with a middle layer dimension of 1920, converting the gradients of LLMs4Code on $(X,Y, X_u, f(X_u))$ and keys with regard to $(X,Y)$ into parameter shifts. 
When training this hypernetwork, the learning rate is set to $1\times10^{-5}$ with 1,000 max steps, and the batch size is 4, while clamping the hypernetwork parameters with $L_2$ norms greater than 1. During editing, the hidden states of the editing instance $(X,Y)$ at the editing layers are first cached, and then the hypernetwork predicts the parameter shifts to complete the weight update.

\textbf{GRACE:} GRACE inserts an adapter into the down projection matrix of the 27th layer of LLMs4Code. Each key’s initial deferral radius is set to 1. GRACE optimizes the values corresponding to the keys using standard fine-tuning with a learning rate of 1 and a maximum of 30 iterations.

Following the established practice in the field of model editing \cite{Meng2022Locating,zhu2020modifyingmemoriestransformermodels,zhang2024comprehensive}, we adopt a strategy where different model editing techniques are directly applied to modify the LLMs4Code for each targeted case, regardless of whether the initial model can produce the correct output.
This decision is based on two reasons: Firstly, as demonstrated in the subsequent section, existing LLMs4Code models are only capable of generating correct outputs for a negligible number of inputs. Therefore, it is reasonable to assume that LLMs4Code does not possess the required code knowledge in advance. Secondly, even if the model already contains the necessary knowledge, our evaluation can still assess the potential side effects generated during the model editing process by utilizing metrics such as Specificity and Fluency.



\section{Study Results}
This section presents details of the study results.

\subsection{RQ1: Proficiency of Editing Techniques on LLMs4Code}

\begin{table*}[t]
\caption{Proficiency of each model editing technique on LLMs4Code.}
\scalebox{0.65}{
\begin{tabular}{ccccccccccc|cccccccccc}
\toprule
Dataset    & \multicolumn{10}{c|}{CNLE (NL2PL)}                     & \multicolumn{10}{c}{CSNE (PL2NL)}                     \\ \hline
Model / Editor &
  \multicolumn{3}{c}{Effectiveness} &
  \multicolumn{3}{c}{Generalization} &
  \multicolumn{3}{c}{Specificity} &
  \multicolumn{1}{c|}{Fluency} &
  \multicolumn{3}{c}{Effectiveness} &
  \multicolumn{3}{c}{Generalization} &
  \multicolumn{3}{c}{Specificity} &
  Fluency \\ \hline
           & EM & BU & RL & EM & BU & RL & EM & BU & RL &  & EM & BU & RL & EM & BU & RL & EM & BU & RL &  \\ \hline
CodeLlama  &  0.29   & 3.23     &  10.39     &  0.02    &3.20      &  10.93     & 94.96      & 99.34      &99.48        &   497.75    &   5.27     & 43.03       & 56.28       & 9.16       &     53.19   & 65.78       & 94.01       & 99.28       & 97.94       &  511.77      \\
FT-L       &  4.79   & 8.73     & 16.57      & 0.53     &4.27      &  12.12     & 68.58      & 93.27      &94.41        &\textbf{ 498.10 }     &    53.56    & 73.22       & 79.84       &  50.37      &  77.26      &  83.79      & 36.34       & 88.89       & 91.01       &  \textbf{511.91}      \\
ROME       &   48.23  &  67.95    &  78.55     & \textbf{37.84}     & \textbf{60.02}     &  \textbf{72.92}     & 0.03      & 1.96      & 9.63       &479.57       &    73.88    & 89.40      &  94.34     &  \textbf{73.88}      & \textbf{89.88}       & \textbf{94.53}       & 0.0       &31.06        & 46.23       & 499.25       \\
MEMIT      &  8.21   &  18.74    &  42.17     & 11.73     &26.63      & 48.49      & 0.0      &   0.67    & 6.64       & 443.87      &  37.31      & 60.79       & 74.71       &  39.32     & 63.75       & 76.77       &  0.0      & 24.0       & 39.3       &  479.03      \\
PMET       &  11.96   &  25.58    &   48.84    &  16.09    & 34.62     &  55.45     & 0.08      &   2.16    & 9.01       &   455.96    &   35.67     &   60.23     &  74.49      & 37.12       &  62.50      & 75.97       &  0.0      &   23.22     &  38.66      & 479.01       \\
MALMEN     &  1.14   & 4.37     & 6.17      & 0.94     &3.64      & 5.31      &  0.62     &  6.02     &  6.99      &440.26       &  12.87      & 17.78       & 19.21       & 14.23       &  19.12      & 20.46       &  0.66      & 15.23       &  17.58      & 498.95       \\
GRACE    &  \textbf{93.67}  & \textbf{96.56}    &  \textbf{97.99}  & 0.03  & 3.2   &10.92   & \textbf{92.89}    &\textbf{99.10}  & \textbf{99.29}  & 476.43 &\textbf{99.63 }  &\textbf{97.76}    & \textbf{99.89}   & 11.46   & 54.10   & 66.48  & \textbf{91.85 }  &  \textbf{99.00}   &\textbf{97.76 }  & 504.57 \\ 
\midrule
CodeQwen   &   0.16  & 5.20     & 17.90      & 0.04     &5.54      & 18.57      &  92.05     &  99.06     & 99.33       & 498.74      &     11.54   & 40.54       & 55.22       & 18.21       & 49.28       &  62.81      &  88.52      & 98.74       &  99.04      &  514.92      \\
FT-L       &  68.40   &  74.05    &  79.05     & \textbf{15.95}     & \textbf{24.22}     &   \textbf{37.31 }   & 14.02      &  62.88     & 69.79  &   \textbf{478.81}   &  \textbf{96.05}     & \textbf{97.38}       & \textbf{99.84}       & \textbf{93.10}       &  \textbf{96.53}      &   \textbf{99.23}     &   0.54     &50.81        &   62.53     &    491.34         \\
ROME       &  3.42  & 10.78     & 30.93      & 3.51     &11.92      & 32.32      &  0.68     &   9.66    & 20.25       & 392.46      & 17.53       & 46.21       &  61.48      & 18.21       &  47.82      &  63.22      & 0.01       &  25.51      &  41.83      &   443.11     \\
MEMIT      & 4.31    &  15.41    &  35.10     & 4.75     & 16.55     &   35.25    &  0.20     & 15.64      & 27.74       & 423.85      &  8.76      &   34.76     & 51.85       & 10.31       &   37.67     & 54.74       &  0.01      & 29.04       & 44.65       & 445.97       \\
PMET       &  1.16   &   8.28   &   25.18    &  1.31    &9.38      & 25.54      &  2.91     &    34.25   & 44.05       & 415.44      &     9.96   &   36.36     & 51.41       & 13.98       & 41.90       &  56.64      & 1.31       & 56.92       & 67.63       & 483.69       \\
MALMEN     &   0.85  &   3.22   &    5.73   &  0.71    & 3.05     &  5.64     &  0.06     &  2.67     & 4.38       &  372.04     &    6.44    &  14.23     &  16.72      &7.10        & 14.78       &  17.22      & 0.0       & 6.66       &  10.38      &  420.12      \\
GRACE   &  \textbf{90.34}  &\textbf{96.03}     &  \textbf{98.03}   & 0.04 & 5.52   &18.53   & \textbf{89.07}   &\textbf{98.73}  & \textbf{99.11}  & 466.86 &  95.79  &97.31    &99.81    &19.29   &49.96   &63.36   & \textbf{86.61 } &\textbf{98.51}   &\textbf{98.87}    &\textbf{503.34}  \\
\midrule
StableCode &   0.18  &  7.82    &  24.25     &  0.02    &9.41      &   26.82    &   85.08    &   98.07    & 98.15       & 495.41      &  0.08      & 24.88       & 40.19       & 0.12       &  29.97      &   45.40     &  86.14      &  98.39      &  98.68      &  519.25      \\
FT-L       &   18.48  & 29.78     & 43.69      &  4.17    & 17.47     & 34.13      & 26.64      & 79.58      & 83.95       &  \textbf{495.17}     &  61.48      & 76.46       &82.99        & 42.77       &  67.41      &  75.83      &33.96        &  85.11      &   88.54     &  \textbf{517.24}      \\
ROME       &   41.93  &  60.84    &  72.05     & \textbf{31.80}     &  \textbf{52.25}    &  \textbf{66.07 }    &  0.0     &    6.97   & 20.84       & 483.78      &   60.07     &  82.23      &  87.65      & \textbf{61.37}       & \textbf{83.18 }      &   \textbf{88.52 }    & 0.0       &  20.75      &  35.70      &  506.73      \\
MEMIT      &  10.60   & 23.71     & 43.75      & 12.75     & 28.39     & 46.90      & 0.0      &  4.38     &  16.51      &  451.48     &    37.07    &    61.83    &  71.26      & 37.55       &  63.12      &  72.32      & 0.0       &  13.72      &    27.83    & 474.10       \\
PMET       &  25.22   &  43.50    & 60.10      & 25.82     &  46.07    &  61.89     &  0.0     & 5.47      &   18.78     & 469.55      &  40.35      & 65.91       &74.49        & 40.97       & 67.21       & 75.69       &0.0        &  15.66      &  29.61      &  483.03      \\
MALMEN     &  0.50   &  3.73    &  6.67     &  0.24    &3.33     &  6.46     &  0.01     & 2.87      & 4.96       &  421.20     & 7.15       & 13.42       & 15.81       &7.93        & 14.38       & 16.70       &  0.0      &5.63        & 9.66       &  465.46      \\
GRACE    &  \textbf{80.09}  &\textbf{91.59}    &\textbf{95.08}    &0.02  & 9.45  &26.88   & \textbf{91.69} & \textbf{97.63} &\textbf{ 97.82} & 483.69 &  \textbf{87.93} & \textbf{95.94}   & \textbf{98.92}   &1.99    & 31.31  & 46.45  & \textbf{83.73 }  &\textbf{98.04}    & \textbf{98.42}   &510.89  \\ 
\bottomrule
\end{tabular}
}
\label{tab:clme_main_res}
\end{table*}

\begin{table}
\caption{The effectivenes decreases from CSNE to CNLE.}
\scalebox{0.75}{
\begin{tabular}{ccccccc}
\hline
Editor     & FT-L    & ROME    & MEMIT   & PMET    & MALMEN  & GRACE   \\ \hline
CodeLlama  & 91.06\% & 34.71\%  & 77.99\% & 66.47\%   & 91.14\%   & 5.98\%  \\
CodeQwen   & 28.79\% & 80.49\% & 50.80\% & 88.35\%   & 86.80\%   & 5.69\%  \\
StableCode & 69.94\% & 30.20\% & 71.41\% & 37.50\% & 93.01\% & 8.92\% \\ \hline
Avg. & 63.26\% & 48.46\% & 67.73\% & 64.11\% &90.31\% & 6.86\% \\ \hline
\end{tabular}}
\label{tab:effectiveness_decreases}
\end{table}

Table \ref{tab:clme_main_res} shows the effectiveness of the six model editing techniques on the CLMEEval benchmark across three LLMs4Code. 
We also list the performances of the original LLMs4Code for convenience.
Overall, we have the following major findings:
\begin{enumerate}
  \item None of the existing model editing techniques can achieve satisfactory results simultaneously across the effectiveness, generalization, and specificity. 
GRACE achieves the best overall performance among all the techniques with the highest effectiveness and specificity values. Specifically, its ROUGE-L value can reach nearly one hundred pencentage (i.e., 99.29\%) when editing CodeLlama on the NL2PL task. Nonetheless, it shows nearly no improvement in generalization compared to the original model.
FT-L achieves good effectiveness and generalization in editing CodeQwen on the CSNE dataset, but its significant decline in specificity and fluency compared to the original model indicates a substantial impact on the non-targeted knowledge and generative capability of the model.
ROME exhibits the best performance in editing CodeLlama on the CSNE dataset, but similarly, it significantly impacts the original knowledge of the model.
The performances of MEMIT, PMET, and MALMEN are generally poor. On the one hand, they do not effectively incorporate the required knowledge into LLMs4Code, which is revealed by the comparatively low effectiveness.
On the other hand, they damage the original knowledge of the model significantly according to the low specificity. 
In terms of fluency, FT-L produces the overall best results, with sometimes even increasing the fluency of the edited model (e.g., when editing CodeLlama on the CSNE dataset).
  \item Most model editing techniques perform poorly on LLMs4Code, being far less proficient compared with editing general LLMs\cite{zhang2024comprehensive}.
For instance, FT-L achieved 100\% effectiveness in terms of the EM when editing Llama-2 on factual knowledge datasets\cite{zhang2024comprehensive}, while its effectiveness on both CNLE and CSNE datasets are generally poor, often being less than 70\%.
ROME, MEMIT, PMET, and MALMEN achieved nearly 100\% effectiveness and over 85\% generalization without significant decreases in specificity and fluency when editing GPT-J\cite{wang2021gpt} on factual knowledge datasets\cite{li2023pmet}. 
However, their effectiveness and generalization on the CNLE and CSNE datasets did not surpass 80\%. Additionally, there was a notable decrease in both specificity and fluency when utilizing these model editing techniques. 
Such results indicate that compared with factual knowledge, editing code-related knowledge is a more challenging task that deserves further investigations in the future.
  \item Generalization is a universal challenge for state-of-the-art model editing techniques. 
Except for four instances where FT-L and ROME edit LLMs4Code on the CSNE dataset, achieving a generalization rate surpassing 50\%, the generalization rate in all other scenarios dips below 50\%.
\item Most editing techniques exhibit significant performance decreases across two different datasets on the same model, with all editing techniques performing worse in NL2PL editing than in PL2NL editing. This suggests that NL2PL editing is a more challenging task even its length of editing target is smaller (cf. Table \ref{tab:data-statistics}). This observation suggests that the difficulty of editing is task-specific, and the presence of more tokens in the editing target does not necessarily indicate a higher level of difficulty.
Table \ref{tab:effectiveness_decreases} reports the percentage decrease in effectiveness based on the exact match metric from CSNE to CNLE for each editing technique. We find that except for GRACE, the performance of other editing techniques decreases by more than 48\% in average, indicating their inability to switch smoothly between different editing tasks. This suggests that GRACE exhibits potential for utilization in various software engineering tasks due to its higher applicability compared to other model editing techniques. 
\end{enumerate}

\begin{figure}[t]
        \centering
        \includegraphics[width=0.45\textwidth]{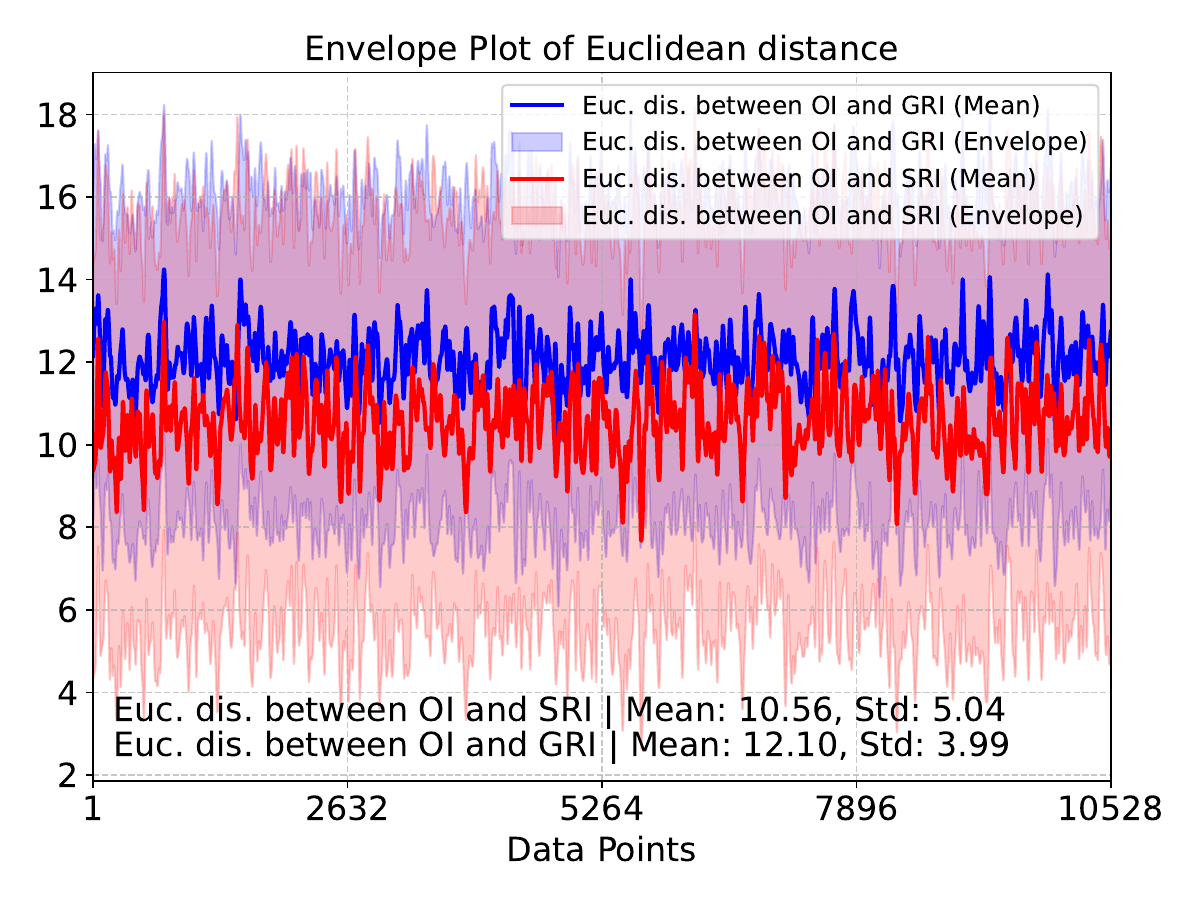}
    \caption{Envelope plot of Euclidean Distance between OI-GRI and that between OI-SRI (keys generated by GRACE).} 
    \label{fig:case_analyze_all}
\end{figure}
Our systematic study of the proficiency of existing
editing techniques on LLMs4Code reveals that:
\begin{tcolorbox}[colback=gray!10,title=,sharp corners=all]
\textit{(1) None of the existing model editing techniques can achieve satisfactory results simultaneously across the effectiveness, generalization, and specificity; (2) Most model editing techniques perform poorly on LLMs4Code, being far less proficient compared with editing general LLMs; (3) Generalization is a universal challenge for state-of-the-art model editing techniques; (4) Most editing techniques exhibit significant performance decreases across two different datasets on the same model, with all editing techniques performing worse in NL2PL editing than in PL2NL editing.}
\end{tcolorbox}

{\bf In-depth Analysis.}
We note that GRACE has the best overall performance among all the techniques, but it performs unsatisfactorily in generalization. We conduct an analysis to delve into the reasons behind its poor generalization.
For convenience, we select the CNLE dataset, which is generally more challenging for existing model editing techniques, and CodeLlama as the subjects for the analysis. 
The study results can naturally be extended to other models and datasets.

The core working mechanism of GRACE involves storing key-value pairs of editing instances in a codebook. A deferral mechanism is used to measure the Euclidean distance between each new input and the keys in the codebook, matching whether the key of the new input falls within the deferral radius of existing keys. If so, the stored values will be used to generate output for this input.
Here, the keys are the last token representations of the inputs, and the values are vectors optimized to produce the desired output from the model. 
An ideal situation would be that the distance between the key of the edited instance and the key of the semantically identical input is relatively close to ensure a good generalization (because they will obtain the identical output). 
Additionally, the distance between the key of the edited instance and the key of the semantically unrelated input should be far apart to ensure a good specificity.
The good specificity of GRACE indicates that the key of edited instance may be far apart to those of unrelated inputs.
However, the poor generalization suggests that the key of edited instance may not be close enough to those of semantically identical inputs.   
As a result, we are motivated to seek for the reasons for the poor generalization of GRACE by investigating the distances among the keys (i.e., last token representations) of different inputs.

Specifically, we denote the OI (Original Intent) as the intent from the CNLE dataset, GRI (Generalization-Related Intent) as the input for testing generalization in CNLE; and SRI (Specificity-Related Intent) as the first specificity case in the CNLE dataset. We input the OI, GRI, and SRI of the test samples into CodeLlama and extract the last token representations, after which we compute the distances between the representations of OI and GRI, and those between the OI and SRI. 
Figure \ref{fig:case_analyze_all} indicates that the distances between the representations of OI and GRI and those between the OI and SRI are severely intertwined, making it difficult to distinguish them using a fixed distance radius. 
In such circumstances, a GRI input might be misclassified as an SRI input by the codebook. Consequently, the edited model may generate incorrect outputs, leading to the poor generalization of GRACE.
To sum up, our analysis reveals that {\bf utilizing the last token representation of inputs as keys is not effective in distinguishing between GRI and SRI}, based on which we design our editing strategy in the following contents.

\subsection{RQ2: Efficiency of Editing Techniques on LLMs4Code}
\begin{figure}[t]
    \centering
    \includegraphics[width=1\linewidth]{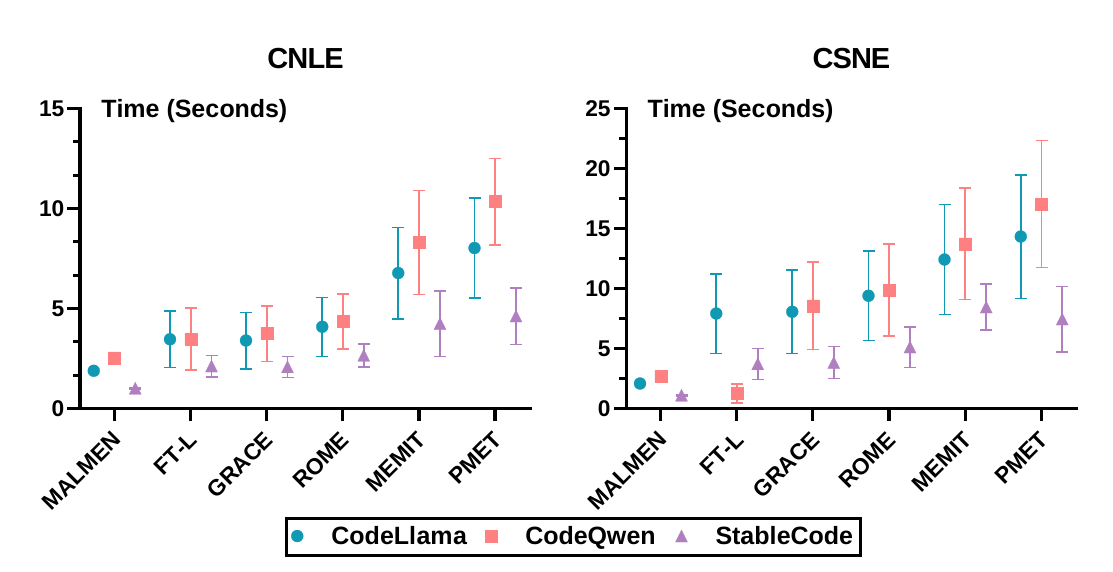}
    \caption{Average time cost of selected model editing techniques per edit.}
    \label{fig:time_efficiency}
\end{figure}
\begin{figure}[t]
    \centering
    \includegraphics[width=1\linewidth]{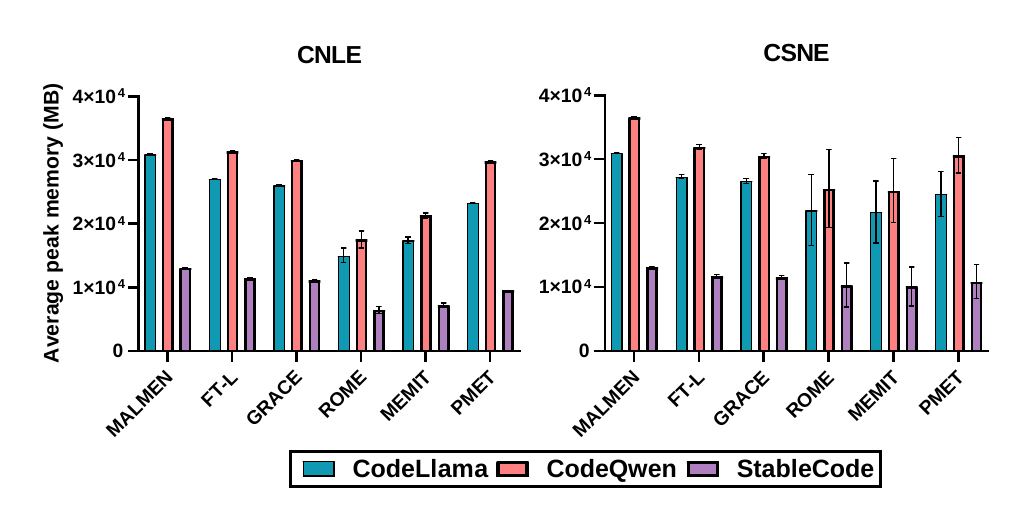}
    \caption{Average peak memory cost of selected model editing techniques per edit.}
    \label{fig:peak_memory_efficiency}
\end{figure}

Figure \ref{fig:time_efficiency} shows the average time cost for each edit using the selected editing techniques. MALMEN takes the least amount of time per edit, completing each edit in approximately two seconds on average. 
PMET and MEMIT have the highest time overhead among these methods, but this does not translate into a performance advantage. In contrast, ROME, FT-L, and GRACE take a moderate time per edit and achieve better overall performance than MALMEN, PMET, and MEMIT.
This indicates that certain model editing techniques can excel in both performance and efficiency. Additionally, the figure shows that the overall editing time for CSNE is slightly longer than that for CNLE. According to the statistics presented in Section \ref{sec:Dataset_statistics}, it is evident that editing longer inputs typically requires more time. Irrespective of dataset and technique variances, there are also differences in the time required for editing different LLMsCode models due to variations in the dimension of the weights. 
CodeQwen has a larger down projection matrix dimension compared to CodeLlama and StableCode, resulting in a longer editing time.

Figure \ref{fig:peak_memory_efficiency} shows the average peak memory on GPU for each edit using these editing techniques. It shows that different models and datasets lead to different peak memory usages, corresponding to the aforementioned editing time. For instance, on both datasets, CodeQwen has the longest per editing time across all editing techniques, and it also has the highest average peak memory. 
Additionally, there is no significant difference in peak memory usage among the different editing techniques. For example, when editing CodeLlama on the CSNE dataset, both the most and least efficient approaches, i.e., ROME and MALMEN, occupy around 30G GPU memory.

In rare cases, MALMEN take around 40GB of memory when editing CodeQwen. This means that model editing techniques, as an efficient knowledge update way, are typically able to update the knowledge of a 7B LLMs4Code on a GPU with 48GB of memory. Our systematic study of the efficiency of existing editing techniques on LLMs4Code reveals that: \begin{tcolorbox}[colback=gray!10,title=,sharp corners=all]
\textit{(1) MALMEN is the most efficient editing technique, while other techniques balance time efficiency with superior performance;
(2) CodeQwen takes longer and uses more memory compared to CodeLlama and StableCode; (3) Despite variations, model editing techniques can update knowledge on large models like those with 7 billion parameters using GPUs within 48GB memory.}
\end{tcolorbox}

\subsection{RQ3: Improvements of Editing Techniques on LLMs4Code}
{\bf Approach.}
To answer this RQ, we seek for the opportunity to improve GRACE. We select GRACE as the study subject since from the above analysis, it achieves the best effectiveness and specificity while maintains high fluency and efficiency.
From the in-depth analysis of RQ1, we find that using the last token representation of the input as the key cannot help GRACE effectively distinguish between GRI and SRI, and thus it currently achieves a comparatively poor generalization.
Therefore, the main objective of this section is to investigate a method to enhance the generalization of GRACE without compromising its performance in other aspects.

The challenge we face is the inability of the last token representation to accurately distinguish the semantic difference between GRI and SRI. 
We are inspired by some recent works that leverage contrastive learning \cite{khosla2020supervised}, a semi-supervised way to automatically capture the key features of the inputs, to boost the performances of diverse software engineering tasks \cite{geng2022fine,cheng2022path,shi2023cocosoda}.
As a result, we decide to exploit contrastive learning in our approach, expecting that the semantic difference between GRI and SRI could be accurately distinguished.

Specifically, we propose \textbf{A-GRACE}, which introduces an MLP encoder composed of two linear layers. 
Different from GRACE, the encoder utilizes the mean of all token representations as the key $q \in \mathbb{R}^{d_n}$, aiming at capturing the information of the entire sequence from a holistic perspective.
The encoding process of the MLP encoder for the key can be described as $q^{*} = W_2\sigma(W_1q)$, where $W_1 \in \mathbb{R}^{d_n \times \frac{d_n}{4}}$, $W_2 \in \mathbb{R}^{\frac{d_n}{4} \times d_o}$, $d_o$ is the output dimension, and $\sigma$ is the activation function. We set $\sigma$ as ReLu.
After that, we train the MLP encoder using contrastive learning on the training set. 
The training target is to differentiate between GRI and SRI, ensuring that the distance between the keys of the OI and the GRI is as close as possible, while the distance between the keys of the OI and the SRI is as far as possible.
To that end, we use the hidden states of the $<$OI, GRI$>$ pairs as positive samples and the hidden states of the $<$OI, SRI$>$ pairs as negative samples for training, with the following loss function \cite{hadsell2006dimensionality}:
\begin{equation}\label{equ:agrace-contrast-learning-loss}
  \mathcal{L} = \frac{1}{N}\sum\limits_{n=1}^{N}{yd^2 +(1-y)\max(margin-d, 0)^2}
\end{equation}
where $N$ is the number of samples, $y$ is the label with 1 for positive samples and 0 for negative samples; $d$ is the Euclidean distance; and $margin$ is a hyperparameter used to control the distance between negative sample pairs.

\begin{table*}[t]
\caption{Proficiency of the A-GRACE approach and its variants on LLMs4Code.}
\scalebox{0.65}{
\begin{tabular}{ccccccccccc|cccccccccc}
\hline
Dataset    & \multicolumn{10}{c|}{CNLE (NL2PL)}                     & \multicolumn{10}{c}{CSNE (PL2NL)}                     \\ \hline
Model / Editor &
  \multicolumn{3}{c}{Effectiveness} &
  \multicolumn{3}{c}{Generalization} &
  \multicolumn{3}{c}{Specificity} &
  \multicolumn{1}{c|}{Fluency} &
  \multicolumn{3}{c}{Effectiveness} &
  \multicolumn{3}{c}{Generalization} &
  \multicolumn{3}{c}{Specificity} &
  Fluency \\ \hline
           & EM & BU & RL & EM & BU & RL & EM & BU & RL &  & EM & BU & RL & EM & BU & RL & EM & BU & RL &  \\ \hline
CodeLlama  &  0.29   & 3.23     &  10.39     &  0.02    &3.20      &  10.93     & 94.96      & 99.34      &99.48        &   497.75    &   5.27     & 43.03       & 56.28       & 9.16       &     53.19   & 65.78       & 94.01       & 99.28       & 97.94       &  511.77      \\
GRACE    &  93.67  & 96.56    &  97.99  & 0.03  & 3.2   &10.92   & 92.89    &99.10  & 99.29  & 476.43 &99.63   &97.76    & 99.89   & 11.46   & 54.10   & 66.48  & 91.85   &  99.00   &97.76   & 504.57 \\ 
\rowcolor[HTML]{CCFFCC} A-GRACE    & 93.57  & 96.56  &97.98    & 64.54  & 73.71  & 78.67  &  37.46 & 73.94 & 79.06 & 459.91 &  99.69 & 97.77   & 99.90    &97.05   & 97.02  & 99.39  & 91.15  & 98.87  & 97.67   & 496.78   \\ 
w/o CL    &  93.67 & 96.58  & 98.00   & 0.03  & 3.21  & 10.92  & 93.10    &99.15  & 99.33 &498.30  & 99.62  &97.75    & 99.89   &26.57   &60.10   &71.19   &91.98   &99.02   & 97.79   & 512.02   \\ 
w/o mean & 93.70  & 96.55  & 98.00   &0.04    &3.23   & 11.01   &92.87   & 99.11   & 99.28   & 477.22   & 99.53  & 97.58   &99.89    &10.02   &53.49   &65.90   &91.69  &98.98   &97.98    & 504.22 \\ 

\hline
CodeQwen   &   0.16  & 5.20     & 17.90      & 0.04     &5.54      & 18.57      &  92.05     &  99.06     & 99.33       & 498.74      &     11.54   & 40.54       & 55.22       & 18.21       & 49.28       &  62.81      &  88.52      & 98.74       &  99.04      &  514.92      \\
GRACE   &  90.34  &96.03     &  98.03   & 0.04 & 5.52   &18.53   & 89.07   &98.73  & 99.11  & 466.86 &  95.79  &97.31    &99.81    &19.29   &49.96   &63.36   & 86.61  &98.51   &98.87    &503.34  \\
\rowcolor[HTML]{CCFFCC} A-GRACE    &  90.52 &  96.01  & 98.02   & 61.42& 73.19  &  79.91 &   39.55 & 78.24 & 82.31 & 446.74 &  95.71 & 97.31  &99.81    & 93.38   &96.55   & 99.32  & 86.29  & 98.47  & 98.84  & 492.55 \\ 
w/o CL    &81.68   & 87.26  & 90.24   & 0.04 & 5.53  &18.53   & 89.07    & 98.73 &99.11  &497.90  & 95.39  &97.13    & 99.60   & 18.83   & 49.55   & 63.00  & 86.82   & 98.59   & 98.93  & 514.85 \\ 
w/o mean    &   90.19&  95.94 & 97.97   &0.03  & 5.49 &18.53   &88.80    &98.70 &99.10  &468.11  & 95.67  & 97.13  & 99.82   &  17.97  &  49.27  & 62.83   & 86.84  &98.62    & 98.97   &503.17  \\ 
\hline
StableCode &   0.18  &  7.82    &  24.25     &  0.02    &9.41      &   26.82    &   85.08    &   98.07    & 98.15       & 495.41      &  0.08      & 24.88       & 40.19       & 0.12       &  29.97      &   45.40     &  86.14      &  98.39      &  98.68      &  519.25      \\
GRACE    &  80.09  &91.59    &95.08    &0.02  & 9.45  &26.88   & 91.69 & 97.63 & 97.82 & 483.69 &  87.93 & 95.94   & 98.92   &1.99    & 31.31  & 46.45  & 83.73   &98.04    & 98.42   &510.89  \\ 
\rowcolor[HTML]{CCFFCC} A-GRACE    & 80.08  & 91.62  &95.10    &57.78 & 73.08  & 80.73  & 33.06    &75.82  &80.75  & 475.07 & 87.85  & 95.88   & 98.86   &85.63    &95.27   &98.44   &83.21   & 97.94  & 98.35  &504.41  \\ 
w/o CL    & 79.85  & 91.50  & 95.01    &11.36  &22.02  &37.60   & 67.94    &83.49  & 85.77 &491.46  & 87.93  &95.97   & 98.93   & 12.75  & 38.60  & 52.43   & 83.86   & 98.10  &98.48   &519.10  \\ 
w/o mean    & 80.00 & 91.67 & 95.11  & 0.02   & 9.59  &27.03    & 81.73   & 97.67    &97.81  & 484.66  &  87.92 &95.68   &98.83   & 0.92   & 30.84   &46.03   & 83.67 & 98.03   & 98.44 &510.31 \\ 

\hline
\end{tabular}
}
\label{tab:agracee_main_res}
\end{table*}

We set the output dimension of the MLP encoder to $d_o=256$ to balance encoding performance and efficiency, and the margin to 1.3 in CNLE and 2.5 in CSNE to enable AGRACE to efficiently distinguish between GRI and SRI.
During the training phase, we use a batch size of 64, a learning rate of 1e-4, and a weight decay of 1e-4 for 100 epochs. We apply an early stopping strategy with a patience of 10 and use the checkpoint with the best performance on the test set as the final MLP encoder weights.

{\bf Results.}
We set the other parameters of A-GRACE to be consistent with those of GRACE. The evaluation results on the CLMEEval benchmark are shown in Table \ref{tab:agracee_main_res}. 
The results show that A-GRACE consistently outperforms GRACE overall, particularly in terms of the generalization. 
For instance, when editing CodeLlama on the CNLE dataset, the EM in terms of the generalization is increased from 0.02\% of the vanilla GRACE to 64.54\% of A-GRACE. On the CSNE dataset, this value is boosted from 9.16\% to 97.05\%.

\begin{figure}[t]
 \centering
        \centering
        \includegraphics[width=0.45\textwidth]{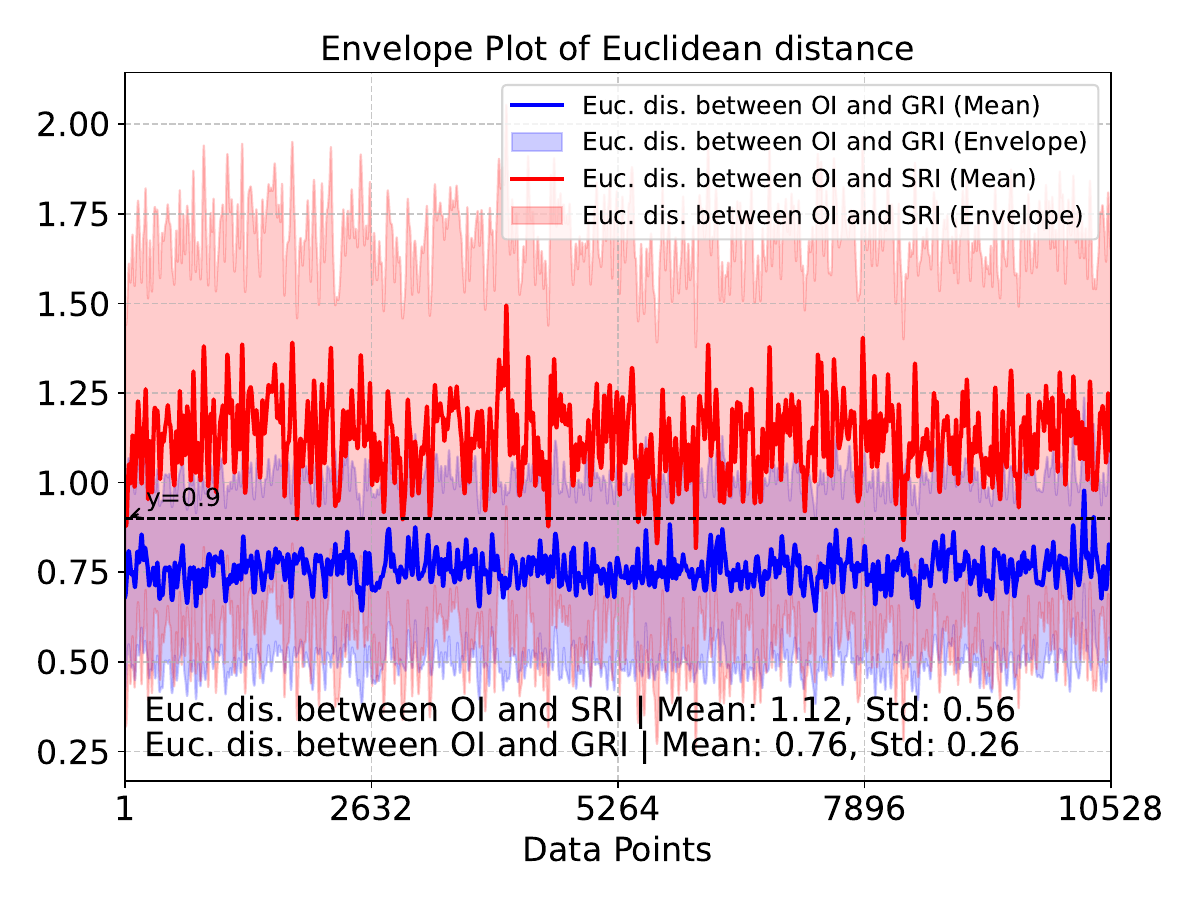}
        \caption{Envelope plot of Euclidean Distance between OI-GRI and that between OI-SRI (keys generated by A-GRACE).}
\label{fig:case_analyze_encode_all}
\end{figure}

\begin{figure}[t]
 \centering
     \begin{subfigure}{0.45\textwidth}
        \centering
        \includegraphics[trim={0cm 0cm 0cm 0cm},clip,width=1\textwidth]{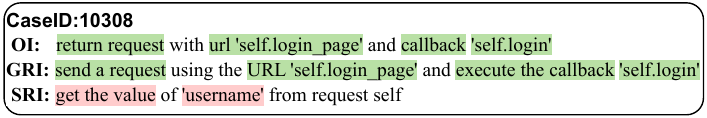}
    \caption{The sampled example.}
    \label{fig:1_case_study}
    \end{subfigure}
    \hfill
    \begin{subfigure}{0.45\textwidth}
        \centering
        \includegraphics[trim={0.9cm 0.7cm 0.7cm 1.75cm},clip,width=\textwidth]{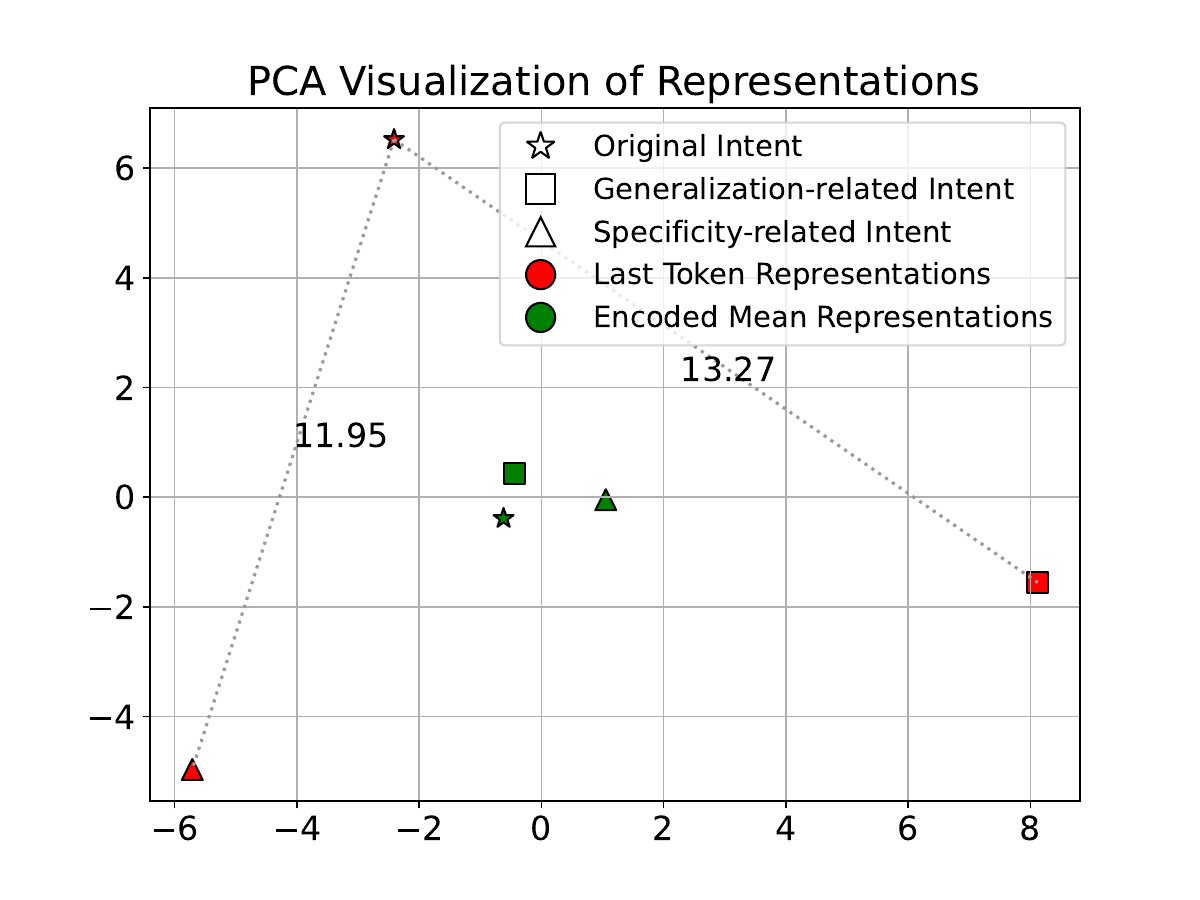}
        \caption{PCA of representations.}
            \label{fig:case_analyze_encode_1_case}
    \end{subfigure}
    \caption{The example and PCA visualization of case analysis.}
\end{figure}

Additionally, in this table, ``w/o CL'' refers to the ablation result of directly using the mean of all token representations as the key without using contrastive learning, and ``w/o mean'' refers to the ablation result of using the MLP encoder trained by contrastive learning to encode the last token representation. 
The ablation results indicate that neither directly using the mean of all token representations nor encoding the last token representation with the MLP encoder trained by contrastive learning can improve generalization. In contrast, AGRACE achieves an order of magnitude improvement in generalization without significant declines in other aspects. Notably, on the CSNE dataset, its performance in other aspects is almost identical to that of GRACE. This demonstrates the rationale of our approach design, that is, using a contrastive-learning-augmented MLP encoder to encode the mean of all token representations as the key is the optimal choice.

In terms of the efficiency, A-GRACE slightly increases the average editing time and memory usage by 4.14\% and 0.38\%, respectively, compare to the vanilla GRACE.
However, the substantial performance gains make this trade-off acceptable.

{\bf Case Analysis.}
We first calculate the distances from the OI to the GRI and SRI for all samples in the CNLE test set, as shown in Figure \ref{fig:case_analyze_encode_all}. 
It can be noted that after encoding, the distance between OI and GRI is significantly smaller than the distance between OI and SRI, which would ease the distinction between the GRI and SRI.
To ease the comprehension, we draw a dotted line which acts as a clear distinction boundary between the blue and the red lines.


To further demonstrate the effect of the MLP encoder, we randomly select an example in Figure \ref{fig:1_case_study}, where we highlight the keywords that lead to the semantic inconsistency between the SRI and OI/GRI.
We input the related intents into CodeLlama to obtain their representations, after which we use PCA to reduce the dimensionality of the last token representations and the mean representations encoded by the MLP encoder to a two-dimensional plane, as shown in Figure \ref{fig:case_analyze_encode_1_case}. 
In the last token representations, the distance between OI and GRI is actually greater than the distance between OI and SRI, indicating that the vanilla GRACE fails to accurately distinguish between GRI and SRI in such a case.
In contrast, in the mean representations encoded by the MLP encoder, the distance between OI and GRI is notably smaller than the distance between OI and SRI.
This enables A-GRACE to recognize that the GRI lies within the deferral radius of the editing target, thereby enhancing its generalization capabilities.
Our systematic study of the improvement of GRACE reveals that: \begin{tcolorbox}[colback=gray!10,title=,sharp corners=all]
\textit{(1) Using the
last token representation as the key cannot help GRACE effectively distinguish between GRI and SRI; (2) By introducing an contrastively-trained encoder in GRACE and encoding the mean of all token representations (the proposed A-GRACE), we successfully enhance the generalization of GRACE without significantly affecting other aspects.
}
\end{tcolorbox}

\section{Threats to Validity}
\textbf{Internal validity.} The data used for testing generalization in the CNLE dataset is rewritten by LLMs based on the original inputs. It is unknown to what extent LLMs can generate semantically identical inputs. To investigate this concern, we randomly sample 100 entries from the CNLE dataset for manual verification.
The first two authors independently assess whether the original requirements are semantically identical to the LLM-revised descriptions, and any conflict is resolved through a discussion.
The results confirm that 95 out of the 100 entries are semantically identical, indicating that the majority of the LLM-generated data, which is used for testing generalization, is reliable.

As shown in Table~\ref{tab:clme_main_res}, modern LLMs achieve less than 1\% exact match on the dataset. There is thus a concern that this could be due to the models not understanding the format of the task and outputting natural language explaining the code alongside the code itself. To better understand this, we randomly selected 50 samples from the CNLE dataset where CodeLlama generates different results compared to the oracle, and manually checked them. Specifically, we investigated whether the answers from CodeLlama contain natural language explanations, and results show that the outputs seldom contain such contents (with only 8 cases being the exceptions). We further confirmed that even if such explanations are removed, the results of CodeLlama still do not match the oracle. 
Moreover, we checked that 45 out of the 50 cases are indeed semantic misalignments between the CodeLlama's outputs and the oracle. 
This indicates that the bias in the evaluation is limited and LLMs4Code are not good at generating completely syntactic-identical outputs.


Our metrics such as BLEU could potentially bring some biases. However, these metrics are commonly-used in the software engineering domain \cite{lin2023cct5,wang2023two,wang2024divide}. These metrics enable the large-scale evaluation of our study. Moreover, they generally keep consistent with the human evaluation as revealed by the recent study \cite{zhou2023codebertscore}.

\textbf{External validity.} Currently, we have focused on two widely-studied SE tasks and the Python language. It is worthwhile to explore more SE tasks and programming languages in the future.

\section{Related Work}
Model editing is a rapidly evolving field, with numerous editing techniques being proposed in recent years for LLMs \cite{zhu2020modifyingmemoriestransformermodels, tan2024massiveeditinglargelanguage, NEURIPS2023_95b6e2ff, Meng2022Locating, meng2022massediting, li2023pmet,mitchell2022fast}. Additionally, many benchmarks for LLMs editing have been introduced to evaluate the performance of these techniques \cite{zhong2023mquake, akyurek2023dune, li2024mikenewbenchmarkfinegrained, li2024unveilingpitfallsknowledgeediting}. These works primarily focus on model editing techniques on general LLMs. The most relevant studies to our work are MENT \cite{gu2024neuronlevel} and CodeUpdateArena \cite{liu2024codeupdatearenabenchmarkingknowledgeediting}. MENT repairs next token errors in code generation by patching specific few neurons in LLMs4Code. CodeUpdateArena is a benchmark that focuses on API function changes for LLMs4Code editing, expecting to address the issue of API function updates through model editing. 

Our work differs from them in several ways. Firstly, we are the first to systematically study the performance of state-of-the-art model editing techniques on LLMs4Code. Secondly, our work focuses on sequence generation tasks (e.g., NL2PL and PL2NL) in LLMs4Code rather than next token prediction, as sequence generation tasks have broader application scenarios in the software engineering domain.
Lastly, based on our evaluation results, we propose A-GRACE, a new model editing approach for LLMs4Code. 

\section{Conlusion}
This paper systematically investigates the performance of six state-of-the-art model editing techniques on LLMs4Code using our proposed CLMEEval benchmark. Through evaluations based on effectiveness, generalization, specificity, and fluency, our study analyzes the strengths and weaknesses of existing editing methods for editing LLMs4Code. Based on the best-performing model editing technique, GRACE, we further propose its augmented version A-GRACE, whose generalization is improved to a large extent. A-GRACE builds a strong baseline for future LLMs4Code editing techniques.

{\bf Source Code.} We have open-sourced all code for this study at: \url{https://github.com/xpq-tech/code-llmedit.git}.

\section{Acknowledgments}
This work was partly supported by the Hunan Provincial Natural Science Foundation Projects (No.2022JJ30668 and No. 2022JJ30046).

\bibliographystyle{IEEEtran}
\bibliography{refs}

\begin{thebibliography}{10}
\providecommand{\url}[1]{#1}
\csname url@samestyle\endcsname
\providecommand{\newblock}{\relax}
\providecommand{\bibinfo}[2]{#2}
\providecommand{\BIBentrySTDinterwordspacing}{\spaceskip=0pt\relax}
\providecommand{\BIBentryALTinterwordstretchfactor}{4}
\providecommand{\BIBentryALTinterwordspacing}{\spaceskip=\fontdimen2\font plus
\BIBentryALTinterwordstretchfactor\fontdimen3\font minus
  \fontdimen4\font\relax}
\providecommand{\BIBforeignlanguage}[2]{{%
\expandafter\ifx\csname l@#1\endcsname\relax
\typeout{** WARNING: IEEEtran.bst: No hyphenation pattern has been}%
\typeout{** loaded for the language `#1'. Using the pattern for}%
\typeout{** the default language instead.}%
\else
\language=\csname l@#1\endcsname
\fi
#2}}
\providecommand{\BIBdecl}{\relax}
\BIBdecl

\bibitem{zhao2023survey}
W.~X. Zhao, K.~Zhou, J.~Li, T.~Tang, X.~Wang, Y.~Hou, Y.~Min, B.~Zhang,
  J.~Zhang, Z.~Dong \emph{et~al.}, ``A survey of large language models,''
  \emph{arXiv preprint arXiv:2303.18223}, 2023.

\bibitem{Chang2024}
\BIBentryALTinterwordspacing
Y.~Chang, X.~Wang, J.~Wang, Y.~Wu, L.~Yang, K.~Zhu, H.~Chen, X.~Yi, C.~Wang,
  Y.~Wang, W.~Ye, Y.~Zhang, Y.~Chang, P.~S. Yu, Q.~Yang, and X.~Xie, ``A survey
  on evaluation of large language models,'' \emph{ACM Trans. Intell. Syst.
  Technol.}, vol.~15, no.~3, mar 2024. [Online]. Available:
  \url{https://doi.org/10.1145/3641289}
\BIBentrySTDinterwordspacing

\bibitem{wei2022emergent}
J.~Wei, Y.~Tay, R.~Bommasani, C.~Raffel, B.~Zoph, S.~Borgeaud, D.~Yogatama,
  M.~Bosma, D.~Zhou, D.~Metzler \emph{et~al.}, ``Emergent abilities of large
  language models,'' \emph{arXiv preprint arXiv:2206.07682}, 2022.

\bibitem{NEURIPS2022_8bb0d291}
T.~Kojima, S.~S. Gu, M.~Reid, Y.~Matsuo, and Y.~Iwasawa, ``Large language
  models are zero-shot reasoners,'' in \emph{Advances in Neural Information
  Processing Systems}, S.~Koyejo, S.~Mohamed, A.~Agarwal, D.~Belgrave, K.~Cho,
  and A.~Oh, Eds., vol.~35.\hskip 1em plus 0.5em minus 0.4em\relax Curran
  Associates, Inc., 2022, pp. 22\,199--22\,213.

\bibitem{wang2024survey}
L.~Wang, C.~Ma, X.~Feng, Z.~Zhang, H.~Yang, J.~Zhang, Z.~Chen, J.~Tang,
  X.~Chen, Y.~Lin \emph{et~al.}, ``A survey on large language model based
  autonomous agents,'' \emph{Frontiers of Computer Science}, vol.~18, no.~6, p.
  186345, 2024.

\bibitem{thirunavukarasu2023large}
A.~J. Thirunavukarasu, D.~S.~J. Ting, K.~Elangovan, L.~Gutierrez, T.~F. Tan,
  and D.~S.~W. Ting, ``Large language models in medicine,'' \emph{Nature
  medicine}, vol.~29, no.~8, pp. 1930--1940, 2023.

\bibitem{Acharya23}
\BIBentryALTinterwordspacing
A.~Acharya, B.~Singh, and N.~Onoe, ``Llm based generation of item-description
  for recommendation system,'' in \emph{Proceedings of the 17th ACM Conference
  on Recommender Systems}, ser. RecSys '23.\hskip 1em plus 0.5em minus
  0.4em\relax New York, NY, USA: Association for Computing Machinery, 2023, p.
  1204–1207. [Online]. Available:
  \url{https://doi.org/10.1145/3604915.3610647}
\BIBentrySTDinterwordspacing

\bibitem{Ali2024}
\BIBentryALTinterwordspacing
A.~Al-Kaswan, ``Towards safe, secure, and usable llms4code,'' in
  \emph{Proceedings of the 2024 IEEE/ACM 46th International Conference on
  Software Engineering: Companion Proceedings}, ser. ICSE-Companion '24.\hskip
  1em plus 0.5em minus 0.4em\relax New York, NY, USA: Association for Computing
  Machinery, 2024, p. 258–260. [Online]. Available:
  \url{https://doi.org/10.1145/3639478.3639803}
\BIBentrySTDinterwordspacing

\bibitem{xu2022systematic}
F.~F. Xu, U.~Alon, G.~Neubig, and V.~J. Hellendoorn, ``A systematic evaluation
  of large language models of code,'' in \emph{Proceedings of the 6th ACM
  SIGPLAN International Symposium on Machine Programming}, 2022, pp. 1--10.

\bibitem{jiang2024survey}
J.~Jiang, F.~Wang, J.~Shen, S.~Kim, and S.~Kim, ``A survey on large language
  models for code generation,'' \emph{arXiv preprint arXiv:2406.00515}, 2024.

\bibitem{wang2023natural}
S.~Wang, M.~Geng, B.~Lin, Z.~Sun, M.~Wen, Y.~Liu, L.~Li, T.~F. Bissyand{\'e},
  and X.~Mao, ``Natural language to code: How far are we?'' in
  \emph{Proceedings of the 31st ACM Joint European Software Engineering
  Conference and Symposium on the Foundations of Software Engineering}, 2023,
  pp. 375--387.

\bibitem{chen2021evaluating}
M.~Chen, J.~Tworek, H.~Jun, Q.~Yuan, H.~P. d.~O. Pinto, J.~Kaplan, H.~Edwards,
  Y.~Burda, N.~Joseph, G.~Brockman \emph{et~al.}, ``Evaluating large language
  models trained on code,'' \emph{arXiv preprint arXiv:2107.03374}, 2021.

\bibitem{geng2024large}
M.~Geng, S.~Wang, D.~Dong, H.~Wang, G.~Li, Z.~Jin, X.~Mao, and X.~Liao, ``Large
  language models are few-shot summarizers: Multi-intent comment generation via
  in-context learning,'' in \emph{Proceedings of the 46th IEEE/ACM
  International Conference on Software Engineering}, 2024, pp. 1--13.

\bibitem{dou2024whatswrongcodegenerated}
\BIBentryALTinterwordspacing
S.~Dou, H.~Jia, S.~Wu, H.~Zheng, W.~Zhou, M.~Wu, M.~Chai, J.~Fan, C.~Huang,
  Y.~Tao, Y.~Liu, E.~Zhou, M.~Zhang, Y.~Zhou, Y.~Wu, R.~Zheng, M.~Wen, R.~Weng,
  J.~Wang, X.~Cai, T.~Gui, X.~Qiu, Q.~Zhang, and X.~Huang, ``What's wrong with
  your code generated by large language models? an extensive study,'' 2024.
  [Online]. Available: \url{https://arxiv.org/abs/2407.06153}
\BIBentrySTDinterwordspacing

\bibitem{lin2022predictive}
B.~Lin, S.~Wang, Z.~Liu, X.~Xia, and X.~Mao, ``Predictive comment updating with
  heuristics and ast-path-based neural learning: A two-phase approach,''
  \emph{IEEE Transactions on Software Engineering}, vol.~49, no.~4, pp.
  1640--1660, 2022.

\bibitem{Ji2023}
\BIBentryALTinterwordspacing
Z.~Ji, N.~Lee, R.~Frieske, T.~Yu, D.~Su, Y.~Xu, E.~Ishii, Y.~J. Bang,
  A.~Madotto, and P.~Fung, ``Survey of hallucination in natural language
  generation,'' \emph{ACM Comput. Surv.}, vol.~55, no.~12, mar 2023. [Online].
  Available: \url{https://doi.org/10.1145/3571730}
\BIBentrySTDinterwordspacing

\bibitem{gu2024neuronlevel}
\BIBentryALTinterwordspacing
J.~Gu, A.~Aleti, C.~Chen, and H.~Zhang, ``Neuron-level llm patching for code
  generation,'' 2024. [Online]. Available:
  \url{https://arxiv.org/abs/2312.05356}
\BIBentrySTDinterwordspacing

\bibitem{tambon2024bugslargelanguagemodels}
\BIBentryALTinterwordspacing
F.~Tambon, A.~M. Dakhel, A.~Nikanjam, F.~Khomh, M.~C. Desmarais, and
  G.~Antoniol, ``Bugs in large language models generated code: An empirical
  study,'' 2024. [Online]. Available: \url{https://arxiv.org/abs/2403.08937}
\BIBentrySTDinterwordspacing

\bibitem{yao-etal-2023-editing}
\BIBentryALTinterwordspacing
Y.~Yao, P.~Wang, B.~Tian, S.~Cheng, Z.~Li, S.~Deng, H.~Chen, and N.~Zhang,
  ``Editing large language models: Problems, methods, and opportunities,'' in
  \emph{Proceedings of the 2023 Conference on Empirical Methods in Natural
  Language Processing}, H.~Bouamor, J.~Pino, and K.~Bali, Eds.\hskip 1em plus
  0.5em minus 0.4em\relax Singapore: Association for Computational Linguistics,
  Dec. 2023, pp. 10\,222--10\,240. [Online]. Available:
  \url{https://aclanthology.org/2023.emnlp-main.632}
\BIBentrySTDinterwordspacing

\bibitem{wang2023knowledge}
\BIBentryALTinterwordspacing
S.~Wang, Y.~Zhu, H.~Liu, Z.~Zheng, C.~Chen, and J.~Li, ``Knowledge editing for
  large language models: A survey,'' 2023. [Online]. Available:
  \url{https://arxiv.org/abs/2310.16218}
\BIBentrySTDinterwordspacing

\bibitem{zhang2024comprehensive}
\BIBentryALTinterwordspacing
N.~Zhang, Y.~Yao, B.~Tian, P.~Wang, S.~Deng, M.~Wang, Z.~Xi, S.~Mao, J.~Zhang,
  Y.~Ni, S.~Cheng, Z.~Xu, X.~Xu, J.-C. Gu, Y.~Jiang, P.~Xie, F.~Huang,
  L.~Liang, Z.~Zhang, X.~Zhu, J.~Zhou, and H.~Chen, ``A comprehensive study of
  knowledge editing for large language models,'' 2024. [Online]. Available:
  \url{https://arxiv.org/abs/2401.01286}
\BIBentrySTDinterwordspacing

\bibitem{meng2022massediting}
\BIBentryALTinterwordspacing
K.~Meng, A.~S. Sharma, A.~Andonian, Y.~Belinkov, and D.~Bau, ``Mass-editing
  memory in a transformer,'' 2022. [Online]. Available:
  \url{http://arxiv.org/abs/2210.07229}
\BIBentrySTDinterwordspacing

\bibitem{mitchell2022fast}
\BIBentryALTinterwordspacing
E.~Mitchell, C.~Lin, A.~Bosselut, C.~Finn, and C.~D. Manning, ``Fast model
  editing at scale,'' in \emph{International Conference on Learning
  Representations}, 2022. [Online]. Available:
  \url{https://openreview.net/pdf?id=0DcZxeWfOPt}
\BIBentrySTDinterwordspacing

\bibitem{li2023pmet}
X.~Li, S.~Li, S.~Song, J.~Yang, J.~Ma, and J.~Yu, ``Pmet: Precise model editing
  in a transformer,'' 2023.

\bibitem{Meng2022Locating}
K.~Meng, D.~Bau, A.~Andonian, and Y.~Belinkov, ``Locating and editing factual
  associations in gpt,'' in \emph{Advances in Neural Information Processing
  Systems}, S.~Koyejo, S.~Mohamed, A.~Agarwal, D.~Belgrave, K.~Cho, and A.~Oh,
  Eds., vol.~35.\hskip 1em plus 0.5em minus 0.4em\relax Curran Associates,
  Inc., 2022, pp. 17\,359--17\,372.

\bibitem{codellama}
B.~Rozière, J.~Gehring, F.~Gloeckle, S.~Sootla, I.~Gat, X.~E. Tan, Y.~Adi,
  J.~Liu, R.~Sauvestre, T.~Remez, J.~Rapin, A.~Kozhevnikov, I.~Evtimov,
  J.~Bitton, M.~Bhatt, C.~C. Ferrer, A.~Grattafiori, W.~Xiong, A.~Défossez,
  J.~Copet, F.~Azhar, H.~Touvron, L.~Martin, N.~Usunier, T.~Scialom, and
  G.~Synnaeve, ``Code llama: Open foundation models for code,'' 2024.

\bibitem{qwen}
J.~Bai, S.~Bai, Y.~Chu, Z.~Cui, K.~Dang, X.~Deng, Y.~Fan, W.~Ge, Y.~Han,
  F.~Huang, B.~Hui, L.~Ji, M.~Li, J.~Lin, R.~Lin, D.~Liu, G.~Liu, C.~Lu, K.~Lu,
  J.~Ma, R.~Men, X.~Ren, X.~Ren, C.~Tan, S.~Tan, J.~Tu, P.~Wang, S.~Wang,
  W.~Wang, S.~Wu, B.~Xu, J.~Xu, A.~Yang, H.~Yang, J.~Yang, S.~Yang, Y.~Yao,
  B.~Yu, H.~Yuan, Z.~Yuan, J.~Zhang, X.~Zhang, Y.~Zhang, Z.~Zhang, C.~Zhou,
  J.~Zhou, X.~Zhou, and T.~Zhu, ``Qwen technical report,'' \emph{arXiv preprint
  arXiv:2309.16609}, 2023.

\bibitem{stable-code-3b}
\BIBentryALTinterwordspacing
N.~Pinnaparaju, R.~Adithyan, D.~Phung, J.~Tow, J.~Baicoianu, and N.~Cooper,
  ``Stable code 3b.'' [Online]. Available:
  \url{[https://huggingface.co/stabilityai/stable-code-3b](https://huggingface.co/stabilityai/stable-code-3b)}
\BIBentrySTDinterwordspacing

\bibitem{NEURIPS2023_95b6e2ff}
T.~Hartvigsen, S.~Sankaranarayanan, H.~Palangi, Y.~Kim, and M.~Ghassemi,
  ``Aging with grace: Lifelong model editing with discrete key-value
  adaptors,'' in \emph{Advances in Neural Information Processing Systems},
  A.~Oh, T.~Naumann, A.~Globerson, K.~Saenko, M.~Hardt, and S.~Levine, Eds.,
  vol.~36.\hskip 1em plus 0.5em minus 0.4em\relax Curran Associates, Inc.,
  2023, pp. 47\,934--47\,959.

\bibitem{tao2012software}
Y.~Tao, Y.~Dang, T.~Xie, D.~Zhang, and S.~Kim, ``How do software engineers
  understand code changes? an exploratory study in industry,'' in
  \emph{Proceedings of the ACM SIGSOFT 20th International symposium on the
  foundations of software engineering}, 2012, pp. 1--11.

\bibitem{ma2024compositional}
Z.~Ma, S.~An, B.~Xie, and Z.~Lin, ``Compositional api recommendation for
  library-oriented code generation,'' in \emph{Proceedings of the 32nd IEEE/ACM
  International Conference on Program Comprehension}, 2024, pp. 87--98.

\bibitem{zhu2020modifyingmemoriestransformermodels}
\BIBentryALTinterwordspacing
C.~Zhu, A.~S. Rawat, M.~Zaheer, S.~Bhojanapalli, D.~Li, F.~Yu, and S.~Kumar,
  ``Modifying memories in transformer models,'' 2020. [Online]. Available:
  \url{https://arxiv.org/abs/2012.00363}
\BIBentrySTDinterwordspacing

\bibitem{tan2024massiveeditinglargelanguage}
\BIBentryALTinterwordspacing
C.~Tan, G.~Zhang, and J.~Fu, ``Massive editing for large language models via
  meta learning,'' 2024. [Online]. Available:
  \url{https://arxiv.org/abs/2311.04661}
\BIBentrySTDinterwordspacing

\bibitem{NEURIPS2020_92650b2e}
J.~Vig, S.~Gehrmann, Y.~Belinkov, S.~Qian, D.~Nevo, Y.~Singer, and S.~Shieber,
  ``Investigating gender bias in language models using causal mediation
  analysis,'' in \emph{Advances in Neural Information Processing Systems},
  H.~Larochelle, M.~Ranzato, R.~Hadsell, M.~Balcan, and H.~Lin, Eds.,
  vol.~33.\hskip 1em plus 0.5em minus 0.4em\relax Curran Associates, Inc.,
  2020, pp. 12\,388--12\,401.

\bibitem{geva-etal-2021-transformer}
\BIBentryALTinterwordspacing
M.~Geva, R.~Schuster, J.~Berant, and O.~Levy, ``Transformer feed-forward layers
  are key-value memories,'' in \emph{Proceedings of the 2021 Conference on
  Empirical Methods in Natural Language Processing}, M.-F. Moens, X.~Huang,
  L.~Specia, and S.~W.-t. Yih, Eds.\hskip 1em plus 0.5em minus 0.4em\relax
  Online and Punta Cana, Dominican Republic: Association for Computational
  Linguistics, Nov. 2021, pp. 5484--5495. [Online]. Available:
  \url{https://aclanthology.org/2021.emnlp-main.446}
\BIBentrySTDinterwordspacing

\bibitem{touvron2023llama2openfoundation}
\BIBentryALTinterwordspacing
H.~Touvron, L.~Martin, K.~Stone, P.~Albert, A.~Almahairi, Y.~Babaei,
  N.~Bashlykov, S.~Batra, P.~Bhargava, S.~Bhosale, D.~Bikel, L.~Blecher, C.~C.
  Ferrer, M.~Chen, G.~Cucurull, D.~Esiobu, J.~Fernandes, J.~Fu, W.~Fu,
  B.~Fuller, C.~Gao, V.~Goswami, N.~Goyal, A.~Hartshorn, S.~Hosseini, R.~Hou,
  H.~Inan, M.~Kardas, V.~Kerkez, M.~Khabsa, I.~Kloumann, A.~Korenev, P.~S.
  Koura, M.-A. Lachaux, T.~Lavril, J.~Lee, D.~Liskovich, Y.~Lu, Y.~Mao,
  X.~Martinet, T.~Mihaylov, P.~Mishra, I.~Molybog, Y.~Nie, A.~Poulton,
  J.~Reizenstein, R.~Rungta, K.~Saladi, A.~Schelten, R.~Silva, E.~M. Smith,
  R.~Subramanian, X.~E. Tan, B.~Tang, R.~Taylor, A.~Williams, J.~X. Kuan,
  P.~Xu, Z.~Yan, I.~Zarov, Y.~Zhang, A.~Fan, M.~Kambadur, S.~Narang,
  A.~Rodriguez, R.~Stojnic, S.~Edunov, and T.~Scialom, ``Llama 2: Open
  foundation and fine-tuned chat models,'' 2023. [Online]. Available:
  \url{https://arxiv.org/abs/2307.09288}
\BIBentrySTDinterwordspacing

\bibitem{conala}
\BIBentryALTinterwordspacing
P.~Yin, B.~Deng, E.~Chen, B.~Vasilescu, and G.~Neubig, ``Learning to mine
  aligned code and natural language pairs from stack overflow,'' in
  \emph{Proceedings of the 15th International Conference on Mining Software
  Repositories}, ser. MSR '18.\hskip 1em plus 0.5em minus 0.4em\relax New York,
  NY, USA: Association for Computing Machinery, 2018, p. 476–486. [Online].
  Available: \url{https://doi.org/10.1145/3196398.3196408}
\BIBentrySTDinterwordspacing

\bibitem{husain2019codesearchnet}
H.~Husain, H.-H. Wu, T.~Gazit, M.~Allamanis, and M.~Brockschmidt,
  ``{CodeSearchNet} challenge: Evaluating the state of semantic code search,''
  \emph{arXiv preprint arXiv:1909.09436}, 2019.

\bibitem{sun2022importance}
Z.~Sun, L.~Li, Y.~Liu, X.~Du, and L.~Li, ``On the importance of building
  high-quality training datasets for neural code search,'' in \emph{Proceedings
  of the 44th International Conference on Software Engineering}, 2022, pp.
  1609--1620.

\bibitem{deepseekv2}
DeepSeek-AI, ``Deepseek-v2: A strong, economical, and efficient
  mixture-of-experts language model,'' 2024.

\bibitem{cohen1960coefficient}
J.~Cohen, ``A coefficient of agreement for nominal scales,'' \emph{Educational
  and psychological measurement}, vol.~20, no.~1, pp. 37--46, 1960.

\bibitem{merrick2024arcticembedscalableefficientaccurate}
\BIBentryALTinterwordspacing
L.~Merrick, D.~Xu, G.~Nuti, and D.~Campos, ``Arctic-embed: Scalable, efficient,
  and accurate text embedding models,'' 2024. [Online]. Available:
  \url{https://arxiv.org/abs/2405.05374}
\BIBentrySTDinterwordspacing

\bibitem{feng-etal-2020-codebert}
\BIBentryALTinterwordspacing
Z.~Feng, D.~Guo, D.~Tang, N.~Duan, X.~Feng, M.~Gong, L.~Shou, B.~Qin, T.~Liu,
  D.~Jiang, and M.~Zhou, ``{C}ode{BERT}: A pre-trained model for programming
  and natural languages,'' in \emph{Findings of the Association for
  Computational Linguistics: EMNLP 2020}, T.~Cohn, Y.~He, and Y.~Liu,
  Eds.\hskip 1em plus 0.5em minus 0.4em\relax Online: Association for
  Computational Linguistics, Nov. 2020, pp. 1536--1547. [Online]. Available:
  \url{https://aclanthology.org/2020.findings-emnlp.139}
\BIBentrySTDinterwordspacing

\bibitem{guo2024exploring}
Q.~Guo, J.~Cao, X.~Xie, S.~Liu, X.~Li, B.~Chen, and X.~Peng, ``Exploring the
  potential of chatgpt in automated code refinement: An empirical study,'' in
  \emph{Proceedings of the 46th IEEE/ACM International Conference on Software
  Engineering}, 2024, pp. 1--13.

\bibitem{yang2022natural}
Z.~Yang, J.~Shi, J.~He, and D.~Lo, ``Natural attack for pre-trained models of
  code,'' in \emph{Proceedings of the 44th International Conference on Software
  Engineering}, 2022, pp. 1482--1493.

\bibitem{BLEU}
\BIBentryALTinterwordspacing
K.~Papineni, S.~Roukos, T.~Ward, and W.-J. Zhu, ``Bleu: a method for automatic
  evaluation of machine translation,'' in \emph{Proceedings of the 40th Annual
  Meeting on Association for Computational Linguistics}, ser. ACL '02.\hskip
  1em plus 0.5em minus 0.4em\relax USA: Association for Computational
  Linguistics, 2002, p. 311–318. [Online]. Available:
  \url{https://doi.org/10.3115/1073083.1073135}
\BIBentrySTDinterwordspacing

\bibitem{ROUGE}
\BIBentryALTinterwordspacing
C.-Y. Lin, ``{ROUGE}: A package for automatic evaluation of summaries,'' in
  \emph{Text Summarization Branches Out}.\hskip 1em plus 0.5em minus
  0.4em\relax Barcelona, Spain: Association for Computational Linguistics, Jul.
  2004, pp. 74--81. [Online]. Available:
  \url{https://aclanthology.org/W04-1013}
\BIBentrySTDinterwordspacing

\bibitem{wan2018improving}
Y.~Wan, Z.~Zhao, M.~Yang, G.~Xu, H.~Ying, J.~Wu, and P.~S. Yu, ``Improving
  automatic source code summarization via deep reinforcement learning,'' in
  \emph{Proceedings of the 33rd ACM/IEEE international conference on automated
  software engineering}, 2018, pp. 397--407.

\bibitem{bansal2021project}
A.~Bansal, S.~Haque, and C.~McMillan, ``Project-level encoding for neural
  source code summarization of subroutines,'' in \emph{2021 IEEE/ACM 29th
  International Conference on Program Comprehension (ICPC)}.\hskip 1em plus
  0.5em minus 0.4em\relax IEEE, 2021, pp. 253--264.

\bibitem{cohen-etal-2024-evaluating}
\BIBentryALTinterwordspacing
R.~Cohen, E.~Biran, O.~Yoran, A.~Globerson, and M.~Geva, ``Evaluating the
  ripple effects of knowledge editing in language models,'' \emph{Transactions
  of the Association for Computational Linguistics}, vol.~12, pp. 283--298,
  2024. [Online]. Available: \url{https://aclanthology.org/2024.tacl-1.16}
\BIBentrySTDinterwordspacing

\bibitem{zhong-etal-2023-mquake}
\BIBentryALTinterwordspacing
Z.~Zhong, Z.~Wu, C.~Manning, C.~Potts, and D.~Chen, ``{MQ}u{AKE}: Assessing
  knowledge editing in language models via multi-hop questions,'' in
  \emph{Proceedings of the 2023 Conference on Empirical Methods in Natural
  Language Processing}, H.~Bouamor, J.~Pino, and K.~Bali, Eds.\hskip 1em plus
  0.5em minus 0.4em\relax Singapore: Association for Computational Linguistics,
  Dec. 2023, pp. 15\,686--15\,702. [Online]. Available:
  \url{https://aclanthology.org/2023.emnlp-main.971}
\BIBentrySTDinterwordspacing

\bibitem{akyurek-etal-2023-dune}
\BIBentryALTinterwordspacing
A.~Aky{\"u}rek, E.~Pan, G.~Kuwanto, and D.~Wijaya, ``{DU}n{E}: Dataset for
  unified editing,'' in \emph{Proceedings of the 2023 Conference on Empirical
  Methods in Natural Language Processing}, H.~Bouamor, J.~Pino, and K.~Bali,
  Eds.\hskip 1em plus 0.5em minus 0.4em\relax Singapore: Association for
  Computational Linguistics, Dec. 2023, pp. 1847--1861. [Online]. Available:
  \url{https://aclanthology.org/2023.emnlp-main.114}
\BIBentrySTDinterwordspacing

\bibitem{wang2024easyediteasytouseknowledgeediting}
\BIBentryALTinterwordspacing
P.~Wang, N.~Zhang, B.~Tian, Z.~Xi, Y.~Yao, Z.~Xu, M.~Wang, S.~Mao, X.~Wang,
  S.~Cheng, K.~Liu, Y.~Ni, G.~Zheng, and H.~Chen, ``Easyedit: An easy-to-use
  knowledge editing framework for large language models,'' 2024. [Online].
  Available: \url{https://arxiv.org/abs/2308.07269}
\BIBentrySTDinterwordspacing

\bibitem{zhang2023large}
Z.~Zhang, M.~Fang, L.~Chen, M.-R. Namazi-Rad, and J.~Wang, ``How do large
  language models capture the ever-changing world knowledge? a review of recent
  advances,'' \emph{arXiv preprint arXiv:2310.07343}, 2023.

\bibitem{li2024sweaupdatingfactualknowledge}
\BIBentryALTinterwordspacing
X.~Li, S.~Li, S.~Song, H.~Liu, B.~Ji, X.~Wang, J.~Ma, J.~Yu, X.~Liu, J.~Wang,
  and W.~Zhang, ``Swea: Updating factual knowledge in large language models via
  subject word embedding altering,'' 2024. [Online]. Available:
  \url{https://arxiv.org/abs/2401.17809}
\BIBentrySTDinterwordspacing

\bibitem{wang2021gpt}
B.~Wang and A.~Komatsuzaki, ``Gpt-j-6b: A 6 billion parameter autoregressive
  language model,'' 2021.

\bibitem{khosla2020supervised}
P.~Khosla, P.~Teterwak, C.~Wang, A.~Sarna, Y.~Tian, P.~Isola, A.~Maschinot,
  C.~Liu, and D.~Krishnan, ``Supervised contrastive learning,'' \emph{Advances
  in neural information processing systems}, vol.~33, pp. 18\,661--18\,673,
  2020.

\bibitem{geng2022fine}
M.~Geng, S.~Wang, D.~Dong, S.~Gu, F.~Peng, W.~Ruan, and X.~Liao, ``Fine-grained
  code-comment semantic interaction analysis,'' in \emph{Proceedings of the
  30th IEEE/ACM International Conference on Program Comprehension}, 2022, pp.
  585--596.

\bibitem{cheng2022path}
X.~Cheng, G.~Zhang, H.~Wang, and Y.~Sui, ``Path-sensitive code embedding via
  contrastive learning for software vulnerability detection,'' in
  \emph{Proceedings of the 31st ACM SIGSOFT International Symposium on Software
  Testing and Analysis}, 2022, pp. 519--531.

\bibitem{shi2023cocosoda}
E.~Shi, Y.~Wang, W.~Gu, L.~Du, H.~Zhang, S.~Han, D.~Zhang, and H.~Sun,
  ``Cocosoda: Effective contrastive learning for code search,'' in \emph{2023
  IEEE/ACM 45th International Conference on Software Engineering (ICSE)}.\hskip
  1em plus 0.5em minus 0.4em\relax IEEE, 2023, pp. 2198--2210.

\bibitem{maas2013rectifier}
A.~L. Maas, A.~Y. Hannun, A.~Y. Ng \emph{et~al.}, ``Rectifier nonlinearities
  improve neural network acoustic models,'' in \emph{Proc. icml}, vol.~30,
  no.~1.\hskip 1em plus 0.5em minus 0.4em\relax Atlanta, GA, 2013, p.~3.

\bibitem{hadsell2006dimensionality}
R.~Hadsell, S.~Chopra, and Y.~LeCun, ``Dimensionality reduction by learning an
  invariant mapping,'' in \emph{2006 IEEE computer society conference on
  computer vision and pattern recognition (CVPR'06)}, vol.~2.\hskip 1em plus
  0.5em minus 0.4em\relax IEEE, 2006, pp. 1735--1742.

\bibitem{lin2023cct5}
B.~Lin, S.~Wang, Z.~Liu, Y.~Liu, X.~Xia, and X.~Mao, ``Cct5: A
  code-change-oriented pre-trained model,'' in \emph{Proceedings of the 31st
  ACM Joint European Software Engineering Conference and Symposium on the
  Foundations of Software Engineering}, 2023, pp. 1509--1521.

\bibitem{wang2023two}
S.~Wang, B.~Lin, Z.~Sun, M.~Wen, Y.~Liu, Y.~Lei, and X.~Mao, ``Two birds with
  one stone: Boosting code generation and code search via a generative
  adversarial network,'' \emph{Proceedings of the ACM on Programming
  Languages}, vol.~7, no. OOPSLA2, pp. 486--515, 2023.

\bibitem{wang2024divide}
S.~Wang, B.~Lin, L.~Chen, and X.~Mao, ``Divide-and-conquer: Automating code
  revisions via localization-and-revision,'' \emph{ACM Transactions on Software
  Engineering and Methodology}, 2024.

\bibitem{zhou2023codebertscore}
S.~Zhou, U.~Alon, S.~Agarwal, and G.~Neubig, ``Codebertscore: Evaluating code
  generation with pretrained models of code,'' in \emph{Proceedings of the 2023
  Conference on Empirical Methods in Natural Language Processing}, 2023, pp.
  13\,921--13\,937.

\bibitem{zhong2023mquake}
Z.~Zhong, Z.~Wu, C.~D. Manning, C.~Potts, and D.~Chen, ``Mquake: Assessing
  knowledge editing in language models via multi-hop questions,'' in
  \emph{Proceedings of the 2023 Conference on Empirical Methods in Natural
  Language Processing}, 2023, pp. 15\,686--15\,702.

\bibitem{akyurek2023dune}
A.~Aky{\"u}rek, E.~Pan, G.~Kuwanto, and D.~Wijaya, ``Dune: Dataset for unified
  editing,'' in \emph{Proceedings of the 2023 Conference on Empirical Methods
  in Natural Language Processing}, 2023, pp. 1847--1861.

\bibitem{li2024mikenewbenchmarkfinegrained}
\BIBentryALTinterwordspacing
J.~Li, M.~Du, C.~Zhang, Y.~Chen, N.~Hu, G.~Qi, H.~Jiang, S.~Cheng, and B.~Tian,
  ``Mike: A new benchmark for fine-grained multimodal entity knowledge
  editing,'' 2024. [Online]. Available: \url{https://arxiv.org/abs/2402.14835}
\BIBentrySTDinterwordspacing

\bibitem{li2024unveilingpitfallsknowledgeediting}
\BIBentryALTinterwordspacing
Z.~Li, N.~Zhang, Y.~Yao, M.~Wang, X.~Chen, and H.~Chen, ``Unveiling the
  pitfalls of knowledge editing for large language models,'' 2024. [Online].
  Available: \url{https://arxiv.org/abs/2310.02129}
\BIBentrySTDinterwordspacing

\bibitem{liu2024codeupdatearenabenchmarkingknowledgeediting}
\BIBentryALTinterwordspacing
Z.~L. Liu, S.~Pandit, X.~Ye, E.~Choi, and G.~Durrett, ``Codeupdatearena:
  Benchmarking knowledge editing on api updates,'' 2024. [Online]. Available:
  \url{https://arxiv.org/abs/2407.06249}
\BIBentrySTDinterwordspacing

\end{thebibliography}
\end{document}